\documentclass[a4paper,11pt]{article}

\usepackage{jheppub} 

\usepackage[T1]{fontenc} 
\usepackage[nolist,printonlyused,smaller]{acronym}
\usepackage{dsfont}
\usepackage{amsmath, amssymb}
\usepackage{mathrsfs}
\usepackage{threeparttable}
\usepackage{multirow}

\newcommand{\unit}{\mathds{1}}

\newcommand{\bg}{\begin{equation}}
\newcommand{\eg}{\end{equation}}
\newcommand{\spl}[1]{\begin{split}#1\end{split}} 
 
\newcommand{\gl}[1]{\eqref{eq:#1}}
\newcommand{\Gl}[1]{Eq.~\eqref{eq:#1}}

\def\m{_{\mu}}
\def\n{_{\nu}}

\def\N{^{\nu}}

\def\r{_{\rho}}

\def\mn{_{\mu\nu}}
\def\MN{^{\mu\nu}}

\makeatletter

\newcommand*\bigcdot@[2]{\mathbin{\vcenter{\hbox{\scalebox{#2}{$\m@th#1\bullet$}}}}}
\makeatother

\newcommand{\mrm}[1]{\mathrm{#1}}
\newcommand{\lef}{\left}
\newcommand{\ri}{\right}

\newcommand{\I}{\mathrm{i}}

\newcommand{\foh}{\frac{1}{2}}
\newcommand{\Tr}{\operatorname{Tr}}

\newcommand{\D}{\mathrm{d}}
\newcommand{\p}{\partial}
\newcommand{\sgo}{\sqrt{g}}

\newcommand{\sgbo}{\sqrt{\bar{g}}}

\newcommand{\gb}{\bar{g}}

\newcommand{\dd}{\D^d}

\newcommand{\mcF}{{\mathcal{F}}}
\newcommand{\cF}{{\mathcal{F}}}

\newcommand{\cO}{{\mathcal{O}}}

\newcommand{\cR}{{\mathcal{R}}}
\newcommand{\mcR}{{\mathcal{R}}}

\newcommand{\cT}{{\mathcal{T}}}

\newcommand{\mcZ}{{\mathcal{Z}}}

\newcommand{\Db}{\bar{D}}
\newcommand{\Rb}{\bar{R}}

\newcommand{\bi}{\begin{itemize}}
\newcommand{\ei}{\end{itemize}}
\newcommand{\be}{\begin{equation}}
\newcommand{\ee}{\end{equation}}

\title{
On the origin of 
almost-Gaussian scaling \\ in asymptotically safe quantum gravity}

\author[a,1]{Maximilian~Becker,\note{Corresponding author.}}
\author[b]{Alexander~Kurov,}
\author[a]{and~Frank~Saueressig}

\affiliation[a]{Institute for Mathematics, Astrophysics and Particle Physics (IMAPP), Radboud University, Heyendaalseweg 135, 6525 AJ Nijmegen, The Netherlands}
\affiliation[b]{Theory Department, Lebedev Physics Institute, Leninsky Prospect 53, Moscow 119991, Russia}

\emailAdd{M.Becker@hef.ru.nl}
\emailAdd{kurov@td.lpi.ru}
\emailAdd{F.Saueressig@hef.ru.nl}

\abstract{The gravitational asymptotic safety program envisions the high-energy completion of gravity through an interacting renormalization group fixed point, the Reuter fixed point. The predictive power of the construction is encoded in the spectrum of the stability matrix which is obtained from linearizing the renormalization group flow around this fixed point. A key result of the asymptotic safety program is that parts of this spectrum exhibits an almost-Gaussian scaling behavior, entailing that operators which are classically highly UV-irrelevant do not induce new free parameters. In this article, we track down the origin of this  property by contrasting the structure of the stability matrix computed from the Wetterich equation and the Composite Operator equation within the realm of $f(R)$-truncations. We show that the almost-Gaussian scaling is not linked to the classical part of the beta functions. It is a quantum-induced almost-Gaussian scaling originating from the quantum corrections in the flow equation. It relies on a subtle interplay among the analytic structure of the theory's two-point function and the way the Wetterich equation integrates out fluctuation modes. As a byproduct we determine the parts of the eigenmode spectrum which is robust with respect to changing the regularization procedure.}

\begin{document} 
\maketitle
\flushbottom

\section{Introduction}
\label{sec:Int}

A key challenge in constructing a quantum field theory including gravitational interactions is the predictive power. This issue already appears in the perturbative quantization of the Einstein-Hilbert action where the negative mass dimension of Newton's coupling leads to new divergences at each order in the loop expansion \cite{'tHooft:1974bx,Goroff:1985sz,Goroff:1985th, vandeVen:1991gw}. Removing these divergences requires the introduction of an infinite tower of counterterms. Following the observation that ``any quantum field theory is renormalizable once all possible counterterms are taken into account'', this procedure may result in a perfectly viable quantum theory. Since each counterterm introduces a new free parameter, such a construction lacks predictive power though. This underlies the assessment that general relativity is an effective field theory \cite{Donoghue:2017pgk} whose high-energy completion requires new physics.

The gravitational asymptotic safety program, recently reviewed in the book chapters \cite{Saueressig:2023irs,Pawlowski:2023gym,Eichhorn:2022gku,Knorr:2022dsx,Morris:2022btf,Martini:2022sll,Wetterich:2022ncl,Platania:2023srt} and text books  \cite{Percacci:2017fkn,Reuter:2019byg}, offers an elegant way to equip this effective field theory with predictive power. The core of the program is an interacting fixed point of the theory's Wilsonian renormalization group (RG) flow, the Reuter fixed point \cite{Reuter:1996cp}. The asymptotic safety hypothesis then states that this fixed point controls the theory as the coarse-graining scale $k$ is sent to infinity. The fact that this hypothesis supplies predictive power can then be seen as follows. By definition, the dependence of the dimensionless couplings $\{u^n\}$ on the coarse-graining scale is encoded in the beta functions 
\be\label{beta2}
k \p_k u^n(k) = \beta^n(u) \, . 
\ee
At a fixed point $\{u^n_*\}$ one has $\beta^n(u_*) = 0$, for all $n$. The RG flow in the vicinity of the fixed point can then be studied by linearizing the beta functions
\be\label{stabmat}
k \p_k u^n(k) = \sum_m (B_*)^n{}_m (u^m - u^m_*) + O((u^j - u^j_*)^2) \, , \quad 
 (B_*)^n{}_m=B^n{}{}_m(u_*) = \left. \frac{\p \beta^n}{\p u^m} \right|_{u = u_*} \, .  
 \ee
Here $B(u_*)$ is the stability matrix of the fixed point. The solutions of the linearized system are readily obtained in terms of the eigenvalues $\lambda_I$ and corresponding (right) eigenvectors $V_I$ of $B(u_*)$ and read
\be\label{sollin}
u^n(k) = u^n_* + \sum_J \, C_J \, V_J^n \left( \frac{k}{k_0} \right)^{\lambda_J} \, . 
\ee
Here, the $C_J$ are free coefficients specifying a solution and $k_0$ is an arbitrary reference scale. For eigendirections with Re($\lambda_J) < 0$, the RG flow is dragged into the fixed point as $k \rightarrow \infty$. Conversely, eigendirections with Re($\lambda_J) > 0$ repel the flow in this limit. In order to reach the fixed point, the integration constants along the UV-repulsive directions must be set to zero while the ones associated with the UV-attractive directions constitute free parameters. Hence the number of free parameters is determined by the spectrum of the stability matrix and equals the number of eigenvalues $\lambda_I$ with negative real part.
More precisely, the predictive power of asymptotic safety hinges on $\mathrm{spec}(B_*)$, being bounded from below and containing a finite number of eigenvalues with $\mathrm{Re}\,\lambda_I < 0$. Conventionally, $\mathrm{spec}(B_*)$ is given in terms of the stability coefficients, also called the critical exponents, $\theta_I$, defined as $\theta_I \equiv - \lambda_I$. Notably, spec($B_*$) is invariant under a redefinition of the couplings $u^n$ \cite{Reuter:2019byg}. For an asymptotically free theory, it agrees with (minus) the classical mass dimension of the operator. Since  the bare action contains a finite number of relevant and marginal couplings only, predictive power is guaranteed. For the asymptotic safety construction, however, the presence of quantum fluctuations makes this feature highly non-trivial.

Determining the spectrum of $B_*$ for the Reuter fixed point and its generalization to gravity-matter systems has been the subject of a vast body of research, see \cite{Lauscher:2002sq,Benedetti:2009rx,Benedetti:2010nr,Benedetti:2013jk,Dietz:2016gzg,Gies:2016con,Denz:2016qks,Falls:2017lst,Ohta:2018sze,Sen:2021ffc,Knorr:2021slg,Baldazzi:2023pep} for selected references. The primary approach builds on solving the Wetterich equation \cite{Wetterich:1992yh,Morris1994,Reuter:1993kw,Reuter:1996cp}, a formally exact functional renormalization group equation (FRGE) encoding the Wilsonian RG flow of the effective average action. The most comprehensive studies along these lines approximate $\Gamma_k$ by a polynomial or a generic function in the Ricci scalar $R$ and determine the beta functions \eqref{beta2} by projecting the RG flow onto a maximally symmetric spherical background \cite{Codello:2007bd,Machado:2007ea,Codello:2008vh,Benedetti:2012dx,Dietz:2012ic,Dietz:2013sba,Bridle:2013sra,Demmel:2014sga,Demmel:2014hla,Demmel:2015oqa,Falls:2013bv,Falls:2014tra,Falls:2016wsa,Alkofer:2018fxj,Alkofer:2018baq,Ohta:2015efa,Ohta:2015fcu,Falls:2016msz,Mitchell:2021qjr,Eichhorn:2015bna,Gonzalez-Martin:2017gza,DeBrito:2018hur,Kluth:2020bdv,Christiansen:2017bsy}. Building on the functional $f(R)$-truncations  \cite{Machado:2007ea,Codello:2008vh,Benedetti:2012dx,Dietz:2012ic,Dietz:2013sba}, Refs.\ \cite{Falls:2014tra,Falls:2018ylp} considered polynomials up to order $N_{\rm prop} = 70$ and observed that the eigenvalue spectrum of classically highly irrelevant operators essentially agrees with the
classical scaling. This feature has been dubbed ``almost-Gaussian scaling'' and has been supported using various approximations of $\Gamma_k$ also beyond the polynomial $f(R)$-computations \cite{Codello:2008vh,Gies:2016con,Falls:2017lst,Ohta:2018sze,Kluth:2022vnq}. Taken together, these studies suggest that the number of free parameters associated with the Reuter fixed point could be as low as $3$ or $4$.

Complementary to solving the Wetterich equation, the stability matrix of a fixed point can also be approximated with a composite-operator equation (COE) \cite{Pawlowski:2005xe,Pagani:2016dof,Pagani:2016pad}. This strategy has been applied to spacetime volumes and also to operators polynomial in the Ricci scalar \cite{Houthoff:2020zqy,Kurov:2020csd,Martini:2023qkp}.\footnote{The initial idea of the COE is that it gives access to the anomalous scaling dimension of composite operators which do not naturally appear in the effective average action. This property has been exploited to study the anomalous scaling dimension of the geodesic distance \cite{Becker:2018quq,Becker:2019tlf,Becker:2019fhi}.} The $f(R)$-type computations \cite{Houthoff:2020zqy,Kurov:2020csd} did not see any evidence for the almost Gaussian scaling behavior. In this case the addition of higher-order operators opened up new UV-relevant directions which is in variance with the results found by solving the Wetterich equation. 

The goal of our work is to track down the origin of this mismatch. In this way, we are able to pinpoint the structural property of the Wetterich equation which underlies the almost Gaussian scaling of the operator spectrum at the Reuter fixed point. The direct comparison shows that this specific property is absent in the COE. This insight allows to tie the mismatch in the two approaches to the pole structure of the trace part of the graviton propagator on a constantly curved background. Moreover, it reveals that the ``almost-Gaussian scaling'' exhibited by the stability matrix is a genuine quantum phenomenon in the sense that it is not linked to the classical part of the beta functions. This analysis is an important step towards understanding  one of the key properties of the gravitational asymptotic safety program, its predictive power. From a wider perspective, our results are also important when comparing the scaling dimensions computed from the composite operator equation to Monte Carlo simulations of the gravitational path integral within Euclidean Dynamical Triangulations (EDT) \cite{Laiho:2017htj} and Causal Dynamical Triangulations (CDT) \cite{Ambjorn:2004qm,Loll:2019rdj,Brunekreef:2023ljt}.

The remainder of this work is organized as follows. In Sec.\ \ref{sec:master}, we derive a master equation giving the structure of the stability matrix at a generic fixed point. In Sec.\ \ref{sec:results}, we compute the building blocks making up the master equation as they appear from solving the Wetterich equation and the composite operator equation for $f(R)$-type operators on spherically symmetric backgrounds. In Sect.\ \ref{sect.analytic}, we link these findings to the analytic structure of the two-point function and we end with a discussion and our conclusions in Sect.\ \ref{sec:conclusion}. In order to make our work self-contained some technical details on $f(R)$-type truncations are given in Appendix \ref{App.A} and \ref{App.B}.

\section{Building the stability matrix at a RG fixed point}
\label{sec:master}
\subsection{General structure of the beta functions}
The primary tool for investigating RG fixed points in quantum gravity is the effective average action $\Gamma_k$, a scale-dependent generalization of the ordinary effective action $\Gamma$. By construction, $\Gamma_k$ provides a description of the effective dynamics which incorporates the effects of quantum fluctuations with momenta $p^2 \gtrsim k^2$ in its effective vertices. In the context of gravity, $\Gamma_k$ depends on a background metric $\gb_{\mu\nu}$ (which may be left unspecified), metric fluctuations $h_{\mu\nu}$ in this background, and ghost fields $\bar{C}_\mu, C^\nu$, arising from gauge-fixing the theory via the Faddeev-Popov construction. Hereby, the background metric $\gb\mn$ and the fluctuations $h\mn$ combine to the full spacetime metric $g\mn$.
The presence of a fixed, non-fluctuating background is essential to be able to define the coarse-graining scale $k$. The interaction monomials in $\Gamma_k$ are taken to be invariant under (background) diffeomorphisms. Conceptually, the gravitational effective average action may be organized according to
\be\label{effavact}
\begin{split}
\Gamma_k[h,\bar{C},C;\gb] = & \, \bar{\Gamma}_k[g] + \hat{\Gamma}_k[h;\gb] + \Gamma^{\rm ghost}_k[h,\bar{C},C;\gb] \, ,
\end{split}
\ee
where $\bar{\Gamma}_k[g]$ is the ``diagonal part'' of $\Gamma_k$ which depends on $\gb_{\mu\nu}$ and $h_{\mu\nu}$ in the combination of $g_{\mu\nu}$ only, $\hat{\Gamma}_k[h;\gb]$ is the off-diagonal part that also incorporates the gauge-fixing term, and $\Gamma^{\rm ghost}_k[h,\bar{C},C;\gb]$ is the ghost-action. In addition, we may supplement the construction by (potentially non-local) ``composite operators'' $\{{\tilde\cO}_n(k)\}$ coupled to their own sources $\varepsilon_n$, 
\be\label{effavact2}
\begin{split}
\Gamma_k[h,\bar{C},C;\gb;\varepsilon] = & \, \Gamma_k[h,\bar{C},C;\gb] +\sum_n\varepsilon_n \, {\tilde\cO}_n(k) \, .
\end{split}
\ee
This extension coincides with the standard definition \eqref{effavact} for $\varepsilon_n=0$.

The dependence of $\Gamma_k$ on the coarse-graining scale $k$ is governed by the Wetterich equation \cite{Wetterich:1992yh,Morris1994,Reuter:1993kw,Reuter:1996cp},
\be\label{FRGE}
	\frac{\mrm d}{\mrm dt} \Gamma_k = \frac{1}{2} \text{STr} \left[ \left( \Gamma^{(2)}_k + \mathcal{R}_k \right)^{-1} \, \frac{\mrm d}{\mrm dt} \mathcal{R}_k \right]\, . 
\ee
Eq.~\eqref{FRGE} is a functional renormalization group equation (FRGE), in which STr denotes a sum over fluctuation modes (including a sum over all fields that $\Gamma_k$ is a functional of) while $\Gamma_k^{(2)}$ denotes the second functional derivative of $\Gamma_k$ with respect to these fluctuation fields. Moreover, $\cR_k$ is a regulator that equips fluctuations of low momenta with a $k$-dependent mass term, while it vanishes for fluctuations with $p^2 \gg k^2$. Further, we introduced the RG time $t \equiv \ln(k/k_0)$ for notational convenience.

Retaining the composite operators and expanding \eqref{FRGE} to first order in $\varepsilon_i$ gives the composite operator equation \cite{Pagani:2016pad,Pagani:2016dof,Becker:2018quq,Becker:2019tlf,Becker:2019fhi,Houthoff:2020zqy,Kurov:2020csd},
\be\label{COE}
\frac{\mrm d}{\mrm dt} \tilde{\cO}_n(k) = - \frac{1}{2} \text{STr} \left[ \left( \Gamma^{(2)}_k + \mathcal{R}_k \right)^{-1} \, \tilde{\cO}_n^{(2)}(k) \left( \Gamma^{(2)}_k + \mathcal{R}_k \right)^{-1} \, \frac{\mrm d}{\mrm dt} \mathcal{R}_k \right]\, . 
\ee
The main virtue of this equation is that it allows to study the scaling dimension of geometric quantities which typically do not appear in $\Gamma_k$. A prototypical example of the latter can be the geodesic distance between points studied in \cite{Becker:2018quq,Becker:2019tlf}. The use of \eqref{COE} requires information about the propagators $\left( \Gamma^{(2)}_k + \mathcal{R}_k \right)^{-1}$ though. This must be provided as an input, e.g., by solving the Wetterich equation appearing at zeroth order in $\varepsilon_n$.  In the sequel, we will focus on composite operators which could also appear within $\Gamma_k[h,\bar{C},C;\gb]$.

The effective average action lives on the gravitational theory space $\cT$. By definition, this linear space comprises all action functionals that can be constructed from the field content and are invariant under background diffeomorphism symmetry. Introducing a basis $\{\cO_n\}$ on $\cT$, we can expand,\footnote{Two comments are in order. Firstly, one can obviously use different bases, say, $\{\cO_n\}$ and $\{\cO'_n\}$, to expand $\Gamma_k$ and $\tilde\cO_m$, respectively. In this work, however, we are only interested in the case $\cO_n=\cO_n'$. Secondly, the expansions~\eqref{expansion} put a non-trivial constraint towards the basis $\{\cO_n\}$, since our definition of $\Gamma_k$ includes the gauge-fixing action which also depends on the theory's couplings. While one might separate the gauge-fixing action from $\Gamma_k$ in order to be able to choose the basis $\{\cO_n\}$ more freely, here, we refrain from doing so for the sake of convenience. Moreover, in Sec.~\ref{sec:results} below we will employ the physical gauge-fixing condition in which the gauge degrees of freedom do not source the anomalous scaling dimension of $\tilde{O}_n$.}
\be\label{expansion}
\Gamma_k[h;\gb] = \sum_n \, \bar{u}^n(k) \, \cO_n \, , \qquad
\tilde{\cO}_m(k) = \sum_n \bar{Z}_m{}^n(k) \, \cO_n \, .
\ee
Here, $\bar{u}^n(k)$ and $\bar{Z}_m{}^n(k)$ are $k$-dependent, dimensionful couplings. Denoting the mass-dimension of the operators as $[\cO_n] = -d_n$, it is convenient to introduce their dimensionless counterparts as
\be\label{dimless}
u^n(k) \equiv \bar{u}^n(k) k^{-d_n} \, , \qquad 
{\gamma}_m{}^n(k) \equiv k^{d_m-d_n} {\left[ \bar{Z}^{-1} {\p_t} \bar{Z} \right]_m}^n \, .  
\ee
The couplings $u^n$ provide natural coordinates on the theory space, while ${\gamma}_m{}^n(k)$ encodes the anomalous scaling dimensions of the composite operators~\cite{Pagani:2016pad}.

 Plugging the expansion of $\Gamma_k$, Eq.~\eqref{expansion}, into the Wetterich equation~\eqref{FRGE}, matching the coefficients of the basis elements $\cO_n$, converting to dimensionless couplings \eqref{dimless}, and solving the resulting system of equations for $\p_t u^n(k)$, we arrive at the beta functions governing the dependence of the couplings $u^n(k)$ on the coarse graining scale,
\be\label{componentflow}
\p_t u^n(k) = \beta^n(u) \, .
\ee
Analogously, the subsitution of \eqref{expansion} into the COE \eqref{COE} yields
\cite{Pagani:2016pad,Pagani:2016dof,Becker:2018quq,Becker:2019tlf,Becker:2019fhi,Houthoff:2020zqy,Kurov:2020csd},
\be\label{compFRGE2}
\sum_{m}{\gamma_n}^m \, k^{d_m}\cO_m = -\frac{1}{2} \, k^{d_n} \, \text{STr}  \left[ \left( \Gamma^{(2)}_k + \mathcal{R}_k \right)^{-1} \,  \cO_n^{(2)} \, \left( \Gamma^{(2)}_k + \mathcal{R}_k \right)^{-1} \, \frac{\mrm d}{\mrm dt} \mathcal{R}_k \right]\, ,
\ee
In this way, $\gamma_{n}{}^m$ becomes a function of the couplings $u^n(k)$, i.e., $\gamma_{n}{}^m(k)\equiv\gamma_{n}{}^m(u(k))$. Note that $\gamma_n{}^m$ also entails an dependence on the beta functions $\beta^n(u(k))$, due to the presence of $\mrm{d}\mathcal R_k/\mrm{d} t$ under the trace.

The goal of this work is to understand the predictive power of the Reuter fixed point based on abstract properties of two-point correlation functions. To do so, we compare the stability matrix $B_*$ as in eq.~\eqref{stabmat} with scaling dimensions obtained from a composite-operator equation. Generally, one may interpret the eigenvalues of the matrix $-d_n\delta_{n}^m+\gamma_{n}{}^m$ as the full scaling dimensions of the operators $\{\cO_n\}$. Thus, we have \textit{two different methods at hand to evaluate critical exponents} -- on the one hand, the (negative) eigenvalues of the stability matrix $(B_*^{\rm FRGE})^{n}{}_m$ obtained from solving the Wetterich equation, and, on the other hand, the eigenvalues of $(B_*^{\rm COE})_{n}{}^m = -d_n\delta_{n}^m+\gamma_{n}{}^m$ obtained from the composite operator equation.

Let us discuss the structure of the beta functions and the stability matrix in more detail. A critical element in their construction is the regulator $\cR_k$ which is conveniently written as\footnote{More precisely, in practice one will set $\cR_k(u;z)=\sum_i\mcZ^{(i)}(\bar u)\left[P_k^{i}(z)-z^i\right]$, where $P_k(z)=z+R_k(z)$ and the index $i$ runs from 1 to highest power of $\Delta = -\Db^2$ appearing in $\Gamma_k^{(2)}$. For the sake of simplicity and without loss of generality, we will remain with the simplified form~\eqref{Rkfact}.}
\be\label{Rkfact}
\cR_k(\bar{u};z) = \mcZ(\bar{u}) \, R_k(z) \, .
\ee
Here, $\mcZ(\bar{u})$ has a status of a wave function renormalization and carries a tensor structure with respect to internal field indices. It depends on $t$ through the dimensionful couplings $\bar{u}$ only. Furthermore, $R_k$ is the scalar regulator. At the level of the Wetterich equation this $\bar{u}$-dependence is dictated by the choice of regularization procedure and is a generic feature \cite{Codello:2008vh}. At the level of the composite operator equation, there is more freedom since there is no direct relation between $\Gamma_k^{(2)}$ and $\cO_n$ which would fix such a dependence based on the composite operators under consideration. As a result of the decomposition \eqref{Rkfact},  one has
\be\label{ptRk}
\frac{\mrm d}{\mrm dt} \mcR_k(\bar{u};z) = 
\left( \sum_n \frac{\p \mcZ(\bar{u})}{\p \bar{u}^n} \frac{\p}{\p t} \bar{u}^n \right) R_k(z) + \mcZ(\bar{u}) \left( \frac{\p}{\p t} R_k \right) \, . 
\ee

Substituting Eqs.~\eqref{expansion} and \eqref{ptRk} into the Wetterich equation, equating coefficients for the basis element $\cO_n$, and retaining the LHS$=$RHS form yields
\be\label{master1}
\p_t u^n + d_n u^n = \omega^n(u) + \sum_m \tilde{\omega}^n{}_m(u) \left(\p_t u^m + d_m u^m \right) \equiv\hat\omega^n(u;\p_t u) \, . 
\ee
This structure disentangles the classical contributions (LHS) and the contributions of the quantum fluctuations (RHS) to the flow. We may think of the latter as a vector field $\hat\omega^n(u,\p_t u)$ on the ``theory's configuation space''. The vector $\omega^n(u)$ and matrix $\tilde{\omega}^n{}_m(u)$ depend on the dimensionless couplings $u^n$, but not on their derivatives $\p_t u^n$. We also observe that $\tilde{\omega}^n{}_m(u)$ may not depend on all couplings, but only those which appear in the regulator $\cR_k(\bar{u})$. Explicit expressions for $\omega^n(u)$ and  $\tilde{\omega}^n{}_m(u)$  can be given in terms of functional traces
\be\label{LHSRHSTMWetterich}
\begin{split}
\omega^n(u) = & \left. \frac{1}{2} \, k^{-d_n} \, \text{STr} \left[ \left( \Gamma^{(2)}_k + \mathcal{R}_k \right)^{-1} \, \mcZ(\bar{u}^i) \left( \frac{\p}{\p t} R_k \right) \right] \right|_{\cO_n} \, ,\\
\tilde{\omega}^n{}_m(u) = & \left. \frac{1}{2} \, k^{d_m-d_n} \, \text{STr} \left[ \left( \Gamma^{(2)}_k + \mathcal{R}_k \right)^{-1} \, \frac{\p \mcZ(\bar{u}^i)}{\p \bar{u}^m} \, R_k \right] \right|_{\cO_n} \, .
\end{split}
\ee
Here, we used the notation $|_{\cO_n}$ to denote the projection of the trace onto the basis element $\cO_n$ and included factors of $k$ in such a way that the LHS is given by dimensionless quantities. The beta functions \eqref{componentflow} are then obtained by solving \eqref{master1} for $\p_t u^n$:
\be\label{master2}
\beta^n(u) = - d_n u^n + \sum_m \left[ (\unit - \tilde{\omega}(u))^{-1} \right]^n{}_m \, \omega^m(u) \, .
\ee

The beta functions~\eqref{master2} solve Eq.~\eqref{master1}, i.e., they fulfill the identity
\be
\beta^n(u) = - d_n u^n+\hat\omega^n(u,\beta(u)) \, .
\ee
We can use this identity to organize the stability matrix obtained from Eq.~\eqref{master2}. Thus, let us consider the (generalized) stability matrix 
\be\label{Budef}
B^n{}_m(u)=\frac{\p}{\p u^m}\beta^n(u)
\ee
at a generic point along the RG trajectory,
\be\label{B1}
\begin{split}
B^n{}_m(u)&=-D^n{}_m+\frac{\p}{\p u^m}\hat\omega^n(u;\beta(u))\\
&=-D^n{}_m+\frac{\p}{\p u^m}\hat\omega_0^n(u)+\sum_l\left[\left(\frac{\p}{\p u^m}\tilde\omega^n{}_l(u)\right)\beta^l(u)+\tilde\omega^n{}_l(u)B^l{}_m(u)\right] \, .
\end{split}
\ee
Here, we have introduced the matrix of classical scaling dimensions, $D^n_m\equiv d_n\delta^n_m$, and the vector $\hat\omega^n_0(u)\equiv\hat\omega^n(u;0)$. While it might be tempting to interpret the RHS of Eq.~\eqref{B1} as the classical scaling, in form of $-D^n_m$, and its appertaining quantum corrections, in form of $\p_m\hat\omega^n$, one must take care with such interpretations at this stage, since $\p_m\hat\omega^n$ depends on the classical scaling dimensions $d_n$ and the stability matrix $B$ itself. Solving Eq.~\eqref{B1} for $B$, one obviously has
\be\label{B2}
B^n{}_m(u)=\sum_l \left[ (\unit - \tilde{\omega}(u))^{-1} \right]^n{}_l \, \left(-D^l_m+ \frac{\p \hat\omega_0^l(u)}{\partial u^m}  +\sum_{l'} \frac{\p \tilde{\omega}^l{}_{l'}(u)}{\partial u^m} \, \beta^{l'}(u) \right)\, .
\ee

In order to relate this expression for the stability matrix with the anomalous scaling dimensions obtained within the composite-operator formalism, let us re-arrange Eq.~\eqref{B1} one more time. Having in mind that $\hat\omega^n=k^{-d_n}(\text{RHS of Eq.~\eqref{FRGE}}|_{\cO_n})$, it is a straightforward calculation to see that
\be 
\frac{\p}{\p u^m}\hat\omega^n(u;\beta(u))=\gamma_m{}^n(u)+\tilde\gamma_m{}^n(u)+\sum_l\tilde\omega^n{}_l(u)B^l{}_m(u) \, ,
\ee
where
\be\label{Gammadef}
\gamma_m{}^n (u) \equiv  - \frac{1}{2} \, k^{d_m - d_n} \, {\rm STr}\left[ \left( \Gamma^{(2)}_k + \mathcal{R}_k \right)^{-1} {\cO}_m^{(2)} \left( \Gamma^{(2)}_k + \mathcal{R}_k \right)^{-1}\, \left. \frac{\mrm d}{\mrm dt} \mcR_k(\bar{u}(k)) \right] \right|_{\cO_n} \, ,
\ee
is the anomalous-dimension matrix of Eq.~\eqref{compFRGE2} for the family of operators $\{\cO_n\}$,\footnote{Note that Eq.~\eqref{Gammadef} involves the relation $\cO_m^{(2)}=\frac{\p \Gamma_k^{(2)}}{\p \bar{u}^m}$. As often the composite-operator equation~\eqref{Gammadef} is evaluated for the case where $\{\cO_n\}$ does not include gauge-fixing contributions, but where $\Gamma_k$ does, we will denote $\gamma$ as $\gamma^\mrm{COE+GF}$ and/or $\gamma^\mrm{COE}$  to distinguish the results in which the gauge-fixing term is included into $\{\cO_n\}$ or not.}
and the matrix $\tilde\gamma_m{}^n(u)$ contains terms involving the $\bar{u}$-derivative of the regulator $\cR_k$,
\be\label{Gammatdef2}
\begin{split}
\tilde\gamma_m{}^n(u) \equiv  & -\frac{1}{2} \, k^{d_m - d_n} \, {\rm STr}\left[ \left( \Gamma^{(2)}_k + \mathcal{R}_k \right)^{-1} \left( \frac{\p \mcZ(\bar{u})}{\p \bar{u}^m} R_k \right) \left( \Gamma^{(2)}_k + \mathcal{R}_k \right)^{-1}\, \left. \frac{\mrm d}{\mrm dt} \mcR_k(\bar{u}(k)) \right] \right|_{\cO_n} \\
& 
+ \frac{1}{2} \, k^{d_m - d_n} \, {\rm STr}\left[ \left( \Gamma^{(2)}_k + \mathcal{R}_k \right)^{-1} \, \left. \frac{\p}{\p \bar{u}^m} \left(\sum_l\frac{\p\mcZ(\bar u)}{\p\bar u^l}d_l\bar u^l R_k+\mcZ(\bar u)\p_t R_k \right) \right] \right|_{\cO_n} \, .
\end{split}
\ee
In summary, for the stability matrix at the non-Gaussian fixed point (NGFP) we have established the formulae
\be\label{master3}
\begin{split}
	(B_*)^n{}_m&=\sum_l \left[ (\unit - \tilde{\omega}(u_*))^{-1} \right]^n{}_l \, \left(-D^l_m+\left. \frac{\p \hat\omega_0^l(u)}{\partial u^m}\right|_{u=u_*}\right)\\
	&=\sum_l \left[ (\unit - \tilde{\omega}(u_*))^{-1} \right]^n{}_l \, \left.\left(-D^l_m+ \frac{\p \omega^l(u)}{\partial u^m}+\frac{\p}{\p u^m}\left(\sum_{l'}\tilde\omega^l{}_{l'}d_{l'}u^{l'}\right)\right)\right|_{u=u_*}\\
	&=\sum_l \left[ (\unit - \tilde{\omega}(u_*))^{-1} \right]^n{}_l \, \left(-D^l_m+\gamma_m{}^n(u_*)+\tilde\gamma_m{}^n(u_*)\right)\, .
\end{split}
\ee
Note that there is no simple identity expressing $\gamma_m{}^n, \tilde{\gamma}_m{}^n$ in terms of $\omega^n$ and $\tilde \omega^n{}_m$, individually. 
The second line of Eq.~\eqref{master3} should then be seen as a rearrangement of the first line, in terms of the piece given by $\gamma_m{}^n$, appearing in the COE, and contributions associated with the dependence of the regulator on the couplings $\bar{u}$, which have been collected in $\tilde{\gamma}_m{}^n$. Only their sum is related to $\hat\omega_0^n{}_m$, namely at the fixed point via $\p\hat{\omega_0}^n{}/\p u^m(u_*)=\gamma_m{}^n(u_*)+\tilde\gamma_m{}^n(u_*)$.

The formulae~\eqref{master3} are one of the main results of the present work. We stress that if the employed regulator $\cR_k$ in the Wetterich equation~\eqref{FRGE} was $\bar{u}$-\emph{in}dependent (which in \cite{Morris:2022btf} is called a non-adaptive cutoff), one has $\tilde\omega=0=\tilde\gamma$ and the critical exponents of the theory would be given by $(B_*)^n{}_m=-D^n_m+\gamma_m{}^n(u_*)$. For this reasoning, one might naively expect this formula to give a generally ``good'' approximation of the stability matrix,
\be\label{Bapproxgamma}
(B_*)^n{}_m\approx-D^n_m+\gamma_m{}^n(u_*) \, ,
\ee
even in the case in which the regulator is coupling-dependent. In particular, one would expect that ``almost-Gaussian scaling'' of the stability coefficients \cite{Falls:2013bv,Falls:2014tra,Falls:2017lst,Falls:2018ylp} means that $B_*$ is dominated by $-D^n_m$ with $\gamma_m{}^n(u_*)$ providing a small correction. With the examples below in Sec.~\ref{sec:results}, we will illustrate that this is not the case though. 
In other words, the matrices $\tilde\gamma_m{}^n(u_*)$ and especially $\tilde\omega^n{}_m(u_*)$ are crucial for determining the spectrum of the stability matrix $B_*$. This observation is important because in many systems, the critical exponents obtained from $B_*$ constitute physical observables. (Whether this is actually the case for gravity is nevertheless up to debate, due to the complexity of constructing observables.) In non-gravitational settings, it is often customary to calculate the stability matrix for fixed values of the anomalous dimension, cf. \cite{Hellwig:2015woa}, following Eq.~\eqref{Bapproxgamma}. This raises the question about the physical relevance of the critical exponents obtained from $(B_*)^n{}_m$ versus those obtained from $-D^n_m+\gamma_m{}^n(u_*)$, in case there is no qualitative agreement among the two.

 Lastly, we point out that for a fixed point corresponding to an asymptotically free theory, the quantum corrections to the stability matrix vanish ($\gamma = \tilde \gamma = \tilde{\omega} = 0$). In this case the stability coefficients agree with the mass dimensions of the couplings.

\section{Almost-Gaussian scaling within $f(R)$-computations}
\label{sec:results}

In this section, we will apply the general results obtained in Sec.~\ref{sec:master} to the well-studied $f(R)$-truncations \cite{Codello:2007bd,Machado:2007ea,Codello:2008vh,Benedetti:2012dx,Dietz:2012ic,Dietz:2013sba,Bridle:2013sra,Demmel:2014sga,Demmel:2014hla,Demmel:2015oqa,Falls:2013bv,Falls:2014tra,Falls:2016wsa,Alkofer:2018fxj,Alkofer:2018baq,Ohta:2015efa,Ohta:2015fcu,Falls:2016msz,Mitchell:2021qjr,Eichhorn:2015bna,Gonzalez-Martin:2017gza,DeBrito:2018hur,Kluth:2020bdv}. Our analysis uses the standard computational framework employed in these works: all computations are carried out using the background of a maximally symmetric $d$-sphere $S^d$. Moreover, we employ the background approximation where the beta functions and anomalous dimensions are read off from the operators
\be\label{projbasis}
\bar{\cO}_n[\gb] = \int \D^dx \sqrt{\gb} \Rb^n \, , 
\ee
i.e., at zeroth order in the fluctuation fields. The scalar part of the regulator is chosen to be of Litim-type \cite{Litim:2001up}, 
\be\label{Rk-litim}
R_k(z)=(k^2-z)\theta(k^2-z) \, , 
\ee
 where $\Theta(x)$ is the Heaviside step function. We first start with a one-dimensional projection in Sec.\ \ref{sec22} before generalizing to the polynomial $f(R)$-case in Sec.\ \ref{sect32}.

\subsection{Warm-up: the stability of Newton's coupling}
\label{sec22}
We start by illustrating the general structures derived in the previous section based on the background Einstein-Hilbert truncation \cite{Reuter:1996cp,Reuter:2001ag,Lauscher:2001ya,Litim:2003vp,Gies:2015tca}, adapting the derivation given in \cite{Reuter:2019byg}. Thus, we solve the Wetterich equation~\eqref{FRGE} and the COE \eqref{compFRGE2} based on the geometric operators
\be\label{COEans}
\begin{split}
\cO_1^{\rm COE} = & \int \D^dx \sqrt{g} R \, ,  \\ 
\cO_1^{\rm COE+GF} = & \int \D^dx \sqrt{g} R + \frac{1}{2\alpha} \int \D^dx \sqrt{\gb} \gb^{\mu\nu} F\m[h;\bar g]F\n[h;\bar g] \, . 
\end{split}
\ee
These operators coincide with \eqref{projbasis} once the fluctuation fields are set to zero. The gauge-condition is taken of the general form
\be\label{gaugechoice}
F\m[h;\bar g]=\bar D\N h\mn-\frac{1+\beta}{d}\bar D\m h\n\N \, . 
\ee
This subsection chooses harmonic gauge, $\alpha = 1, \beta = (d-2)/2$. 
 The inclusion of the gauge-fixing term ensures that $\cO^{(2)}_1$ depends on the background covariant derivative $\bar{D}_\mu$ in the form of the background Laplacian $\Delta =  - \gb^{\mu\nu} \Db_\mu \Db_\nu$ only.

In the first step, we project the Wetterich equation onto the subspace spanned by $\bar{\cO}_1[\gb]$. Thus, all vectors and matrices associated with this example are one-dimensional. Including the hamonic gauge-fixing condition, the corresponding ansatz for the effective average action reads
\be
\bar{\Gamma}_k[g] + \hat{\Gamma}_k[h;\gb] \simeq \bar{u}^1 \, \cO_1^{\rm COE+GF} \, , \quad\text{with}\quad u^1 \equiv k^{2-d} \, \bar{u}^1 \, .
\ee
The coupling $\bar{u}^1$ is related to Newton's coupling $G_k$ by $\bar{u}^1 = - \frac{1}{16 \pi G_k}$ and we have $[\bar{u}^1] \equiv D_u = d-2 > 0$.  Following the computation  \cite{Reuter:2019byg}, the LHS$=$RHS form of the projected Wetterich equation, introduced in \eqref{master1}, is
\be\label{EH3}
\p_t u^1 +D_u \, u^1 =\omega +\tilde\omega(u^1)\, 
\left( \p_t u^1 +D_u \, u^1 \right) \, .
\ee
The vector $\omega(u)$ and matrix $\tilde{\omega}(u)$ given by  $\omega =   B_1$ and  $\tilde{\omega}(u^1) = B_2/u^1$, with the constants 
\be\label{B1B2}
\begin{split}
B_1 = & \frac{1}{(4\pi)^{d/2}} \, \frac{1}{12} \left( d(d-3) \Phi^1_{d/2-1} - 6 (d^2-d+4) \Phi^2_{d/2} \right) \, , \\
B_2 = & \frac{1}{(4\pi)^{d/2}} \, \frac{1}{24} \, \left( d(d+1) \tilde{\Phi}^1_{d/2-1} - 6 d (d-1) \tilde{\Phi}^2_{d/2} \right) \, . 
\end{split}
\ee
The dimensionless threshold functions $\Phi^p_n(w)$ and $\tilde{\Phi}^p_n(w)$ introduced in \cite{Reuter:1996cp} are evaluated at vanishing argument and carry the dependence of the result on the choice of scalar regulator. For the Litim-type profile \eqref{Rk-litim} they evaluate to
\be
\Phi^p_n=\frac{1}{\Gamma(n+1)} \, , \qquad \tilde{\Phi}^p_n= \frac{1}{\Gamma(n+2)} \, . 
\ee
Notably, the present case is special in the sense that $\omega$ is independent of the coupling $u^1$. 

Solving \eqref{EH3} for $\p_t u^1$ gives the beta function%
\be
\beta_u(u^1)= -D_u\,u^1+\frac{\omega}{1-\tilde\omega(u^1)} \, .
\ee
This beta function admits a single NGFP characterized by
\be
u^1_* = B_2 + \frac{B_1}{D_u} \, , \qquad B_*^{\rm FRGE} = \, - D_u + D_u^2 \, \frac{B_2}{B_1} \, .  
\ee
Evaluating these expressions for $d=4$ yields
\be
u^1_* \simeq -0.012 \, , \qquad B_*^{\rm FRGE}= - 2.09 \, . 
\ee
Thus, we encounter a NGFP situated at a positive Newton's coupling and a UV-attractive eigendirection.

In the next step, we compare this result to the one found in  the composite operator approximation \eqref{Bapproxgamma}. Thus, we seek to find the anomalous scaling dimension of the operator $\bar{\cO}_1[\gb]$. Without the inclusion of the gauge-fixing term in \eqref{COEans}, this computation has been carried out in \cite{Houthoff:2020zqy}, yielding
\be\label{gammaev}
\begin{split}
\gamma^{\rm COE}(u^1_*) = - \frac{1}{(4\pi)^{d/2}} &  \, \frac{1}{u^1_*} \, \Bigg[ 
\frac{1}{24} \left( d^3+9 d^2 - 22 d -24  \right) \, 
\left( \Phi^2_{d/2} - \frac{1}{2} (2-d) \tilde{\Phi}^2_{d/2} \right) \\
& - \frac{1}{2} \left(d^3 - 3d^2 + 4d -8 \right) \left( \Phi^3_{d/2+1} - \frac{1}{2} (2-d) \tilde{\Phi}^3_{d/2+1} \right) 
\Bigg] \, . 
\end{split}
\ee
In comparison to the ``exact definition'' of $\gamma$, Eq.~\eqref{Gammadef}, this result neglects the contribution of the gauge-fixing terms in $\frac{\p \Gamma_k^{(2)}}{\p \bar{u}^m}$. 
Adding the gauge-fixing term (implementing harmonic gauge) changes this result to
\be\label{gammaev2}
\begin{split}
\gamma^{\rm COE+GF}(u^1_*) = - \frac{1}{(4\pi)^{d/2}} & \, \frac{1}{u^1_*} \, \Bigg[  
\frac{1}{24} d \, (d^2+11d-14) \, 
\left( \Phi^2_{d/2} - \frac{1}{2} (2-d) \tilde{\Phi}^2_{d/2} \right) \\
& - \frac{1}{2} d^2(d-1) \left( \Phi^3_{d/2+1} - \frac{1}{2} (2-d) \tilde{\Phi}^3_{d/2+1} \right) 
\Bigg] \, . 
\end{split}
\ee
Using these results in \eqref{Bapproxgamma} and specializing to $d=4$ then yields
\be
B_*^{\rm COE} =-2+\gamma^{\rm COE}(u^1_*)= -1.91 \, , \qquad B_*^{\rm COE+GF} = -2 + \gamma^{\rm COE+GF}(u^1_*) = -1.94 \, . 
\ee
Hence, we again encounter a UV-attractive eigendirection. In both computations $B_*$ is governed by the classical part $-D_u$ and receives a subleading correction from the quantum corrections. We also observe that there is a reasonable qualitative agreement between the exact projection and its approximation by the composite operator contribution. Subsequently, we will observe how this approximation becomes increasingly worse when going to increasing the order of the truncation.

\subsection{Polynomial $f(R)$-truncations}
\label{sect32}
One of the most prominent results in the context of asymptotically safe quantum gravity is the ``almost-Gaussian'' scaling observed when considering polynomial $f(R)$-truncations at sufficiently large order of the polynomials \cite{Falls:2013bv,Falls:2014tra}. This statement paraphrases that the spectrum of critical exponents $\{\theta_I^\mrm{FRGE}=-\lambda_I^\mrm{FRGE}\}$ of $B^\mrm{FRGE}_*$ converges to the Gaussian scaling dimensions $\{d_I\}$. Hereby, the corrections to the first three classical scaling dimensions, $d_0=4$, $d_1=2$, and $d_2=0$ are such that the first two directions remain relevant, whereas the quantum corrections shift the third direction  from a marginal to a relevant direction. All other directions remain irrelevant. For operators $\cO_n$ with sufficiently large value of $n$ this is ensured by the aforementioned convergence pattern. Therewith, the polynomial $f(R)$-truncation exhibits a three-dimensional UV-critical hypersurface.

In this section, we carry out a first reconnoitring study comparing the stability matrices at the fixed point obtained from the standard FRGE procedure for the $f(R)$-truncation and the corresponding composite operator results. The FRGE computation is based on Eq.~\eqref{master3}, i.e., 
\be\label{Bfrge}
\begin{split}
	(B^\mrm{FRGE}_*)^n{}_m&=\sum_l \left[ (\unit - \tilde{\omega}(u_*))^{-1} \right]^n{}_l \, \left.\left(-D^l_m+ \frac{\p \omega^l(u)}{\partial u^m}+\frac{\p}{\p u^m}\left(\sum_{l'}\tilde\omega^l{}_{l'}d_{l'}u^{l'}\right)\right)\right|_{u=u_*}\\
	&=\sum_l \left[ (\unit - \tilde{\omega}(u_*))^{-1} \right]^n{}_l \, \left(-D^l_m+\gamma_m{}^n(u_*)+\tilde\gamma_m{}^n(u_*)\right)\, .
\end{split}
\ee
The corresponding results for the scaling corrections emerging from the COE are obtained from Eq.~\eqref{Bapproxgamma}, i.e.,
 \be\label{Bapproxgamma2}
	(B^\mrm{COE}_*)^n{}_m=-D^n_m+\gamma_m{}^n(u_*)\, .
\ee
This comparison thus boils down to the question whether we may neglect the contributions of the matrices $\tilde\omega^n{}_m(u_*)$ and $\tilde\gamma_m{}^n(u_*)$. If we do so for the contribution of $\tilde\gamma_m{}^n(u_*)$, we may think of the matrix $\tilde\omega^n{}_m(u_*)$ as translating between the two approximations.

We start by analyzing Eq.\ \eqref{Bfrge}. For functional $f(R)$-truncations the ingredients in 
 Eq.~\eqref{effavact} take the form
\be\label{ansatz}
\begin{split}
\bar\Gamma_k[g]&=\int\dd x\,\sqrt{g}f_k(R)=\int\dd x\,\sqrt{g}k^d\mcF_k(\rho) \, ,\\
\hat\Gamma[h;\bar g]&
=\frac{1}{2\alpha}\int\dd x\sqrt{\bar{g}}\, \bar g\MN F\m[h;\bar g]F\n[h;\bar g]\, .
\end{split}
\ee
Here $f_k(R)$ is a generic, $k$-dependent function of the scalar curvature $R$ with dimensionless counterparts, $\mcF_k(\rho)=k^{-d}f_k(R)$ and $\rho=k^{-2}R$. Supplementing \eqref{ansatz} by the corresponding ghost action and substituting the result into the Wetterich equation results in a partial differential equation governing the functional form of $\mcF_k(\rho)$. There are several incarnations of this equation which differ by the treatment of the discrete modes appearing on spherical backgrounds \cite{Morris:2022btf}. 

The technical construction of the flow equation for $f(R)$-truncations implements the York decomposition \cite{York,York2} of the fluctuation fields
\be\label{TTdec}
h_{\mu\nu} = h\mn^\mrm{TT} + \Db_\mu  \xi_\nu + \Db_\nu  \xi_\mu +  \Db_\mu \Db_\nu \sigma - \frac{1}{d} \gb_{\mu\nu} \Db^2 \sigma + \frac{1}{d} \gb_{\mu\nu} h \, .
\ee
Here $h\mn^\mrm{TT}$ is the transverse-traceless (TT) part, $\hat\xi\m$ is the transverse vector field of the longitudinal part of $h\mn$, $\sigma$ is the corresponding longitudinal scalar, and $h$ is the trace part. The component fields are subject to the constraints
\be
\Db^\mu h\mn^\mrm{TT} = 0 \, , \quad \gb^{\mu\nu} h\mn^\mrm{TT} = 0 \, , \quad  \Db_\mu \xi^\mu = 0 \, , \quad \gb^{\mu\nu} h\mn^\mrm{TT} = h \, . 
\ee
The fields are re-scaled to be orthonormal on Einstein spaces,
\be
\hat\xi\m = (2(\Delta - \Rb/d))^{1/2} \xi_\mu \, , \quad \hat\sigma = \left(\frac{d-1}d \Delta \left(\Delta - \Rb/(d-1) \right) \right)^{1/2} \sigma \, , \quad \hat{h} = \frac{1}{\sqrt{d}} h \, , 
\ee
which also accounts for the Jacobians arising from the decomposition \eqref{TTdec}.

Owed to the higher-derivative terms generated by $\bar{\Gamma}_k$ it turns out to be convenient to restrict \eqref{ansatz} to Landau-type gauges, which fix $\beta = 0$ and take the limit $\alpha \rightarrow 0$ once all ingredients in the traces are computed. The ghost-sector arising from this ``physical gauge'' follows \cite{Machado:2007ea}. 
This choice leads to a number of simplifications. Firstly, the matrix $(\Gamma_k^{(2)} + \cR_k)$ becomes diagonal in field space. Secondly, the contributions of the unphysical transverse vector field $\xi\m$ and the unphysical scalar field $\sigma$ become independent of the function $f_k(R)$. This is readily deduced by analyzing the $\alpha$-dependence of the decomposed propagator (see Table 1 in \cite{Machado:2007ea} and Eqs. (97)-(99) of \cite{Codello:2008vh}). Conversely, the matrix elements for the transverse traceless and $\hat{h}$-fluctuations appearing in $(\Gamma_k^{(2)} + \cR_k)$ are solely governed by $\bar{\Gamma}_k[g]$. Specifying to the Litim-type regulator, they read \cite{Codello:2008vh}
\be\label{Gamma2}
\begin{split}
(\Gamma_{k,\mrm{TT}}^{(2)} + \cR_{k,\mrm{TT}}) = & - \frac{1}{2} \Bigg[ \left( k^2 - \frac{2 (d-2)}{d(d-1)} \Rb \right) f^\prime_k + f_k \Bigg] \, \unit_{\mrm{TT}} \, , \\
(\Gamma_{k,\hat{h}\hat{h}}^{(2)} + \cR_{k,\hat{h}\hat{h}}) = & \frac{d-2}{4}\Bigg[  \frac{4(d-1)^2}{d(d-2)}f^{\prime\prime}_k \left( k^2 - \frac{\Rb}{d-1} \right)^2 \\ &  \qquad \qquad + \frac{2(d-1)}{d} f^\prime_k \left( k^2 - \frac{\Rb}{d-1} \right)  - \frac{2 \Rb}{d} f^\prime_k + f_k \Bigg] \, . 
\end{split}
\ee
Here, $\unit_{\mrm{TT}}$ is the identity on the space of transverse-traceless fields, $f'_k(R)=\p_R f_k(R)$, and we adopt an identical notation for the dimensionless functions $\mcF_k'(\rho)=\p\r\mcF_k(\rho)$. The explicit form of corresponding matrix element of the regulator $\cR_k$ arising from implementing the adaptive cutoff $\Delta \mapsto \Delta + R_k$ is
\be\label{cRhh}
\begin{split}
\cR_{k,\mrm{TT}} = & - \frac{1}{2} \, f^\prime_k \, R_k  \, \unit_{\mrm{TT}} \, ,
\\
\cR_{k,\hat{h}\hat{h}} = & \frac{d-1}{2 d} \Big[(d-2) f^{\prime}_k+2 (d-1) f^{\prime\prime}_k (2 \Delta +R_k)-4 f^{\prime\prime}_k \Rb\Big] R_k \, . 
\end{split}
\ee
The Wetterich equation takes the schematic form
\be\label{FRGE2}
\begin{split}
	\frac{\mrm d}{\mrm dt}\bar\Gamma_k=&\,\foh\Tr_\mrm{TT}\left[\left(\bar\Gamma_{k,\mrm{TT}}^{(2)}+\cR_{k,\mrm{TT}}\right)^{-1}\frac{\mrm d}{\mrm dt} \cR_{k,\mrm{TT}}\right]
	+\foh\Tr_\mrm{S}\left[\left(\bar\Gamma_{k,{\hat h\hat h}}^{(2)}+\cR_{k,{\hat h\hat h}}\right)^{-1}\frac{\mrm d}{\mrm dt} \cR_{k,{\hat h\hat h}}\right]\\
	&+\bar\Gamma_k \text{-independent terms} \, ,
\end{split}
\ee
where the $\bar\Gamma_k$-independent terms are due to the Fadeev-Popov ghosts and the contributions of the $\xi_\mu$ and $\sigma$ modes.
Owed to the presence of the higher-derivative terms, the regulator does not obey the simple factorization \eqref{Rkfact}. Nevertheless, one is able to separate the trace contributions into parts associated with the regulator and the couplings appearing within \eqref{cRhh}.
The result of this computation is rather lengthy and given as Eq.\ \eqref{eq:B1} of Appendix \ref{App.B}.

We then project Eq.~\eqref{FRGE2} onto the basis \eqref{projbasis}. Introducing the dimensionless volume $\tilde V=k^d \int\dd x\sgbo\propto\rho^{-d/2}$, the LHS of Eq.~\eqref{FRGE2}  is
\be\label{LHS}
\frac{\mrm d}{\mrm dt}\bar\Gamma_k=\tilde V\left\{ (\p_t\mcF_k)(\rho)+d\,\mcF_k(\rho)-2\rho\mcF_k'(\rho)\right\} \, .
\ee
Since in general the regulator is of the form $\cR_k=\cR_k[f_k',f_k'']$, the RHS can be written as 
\be\label{RHS}
\text{RHS}=\tilde V\left\{I_0[\mcF_k](\rho)+\left(k^{-d+2}\frac{\mrm d}{\mrm dt} f_k'\right)(\rho)I_1[\mcF_k](\rho)+\left(k^{-d+4}\frac{\mrm d}{\mrm dt} f_k''\right)(\rho)I_2[\mcF_k](\rho)\right\} \, .
\ee
Each function $I_0$, $I_1$, and $I_2$ contains contribution from the TT- and trace-sector as well as from the $f_k(R)$-independent terms. For details, we refer the reader to~\cite{Falls:2014tra}. 

The transition from the functional to the polynomial form of 
 Eqs.~\eqref{LHS} and~\eqref{RHS} uses the expansion
\be\label{ansatzfR}
f_k(R)=\sum_{n=0}^{N_\mrm{prop}}\bar u^n(k) R^n\, ,\quad\mcF_k(\rho)=\sum_{n=0}^{N_\mrm{prop}} u^n(k)\rho^n \, ,
\ee
where the polynomials are truncated at $N_{\rm prop}$. The LHS given in Eq.~\eqref{LHS} then becomes
\be 
\frac{\mrm d}{\mrm dt}\bar\Gamma_k=\tilde V \sum_{n=0}^{N_\mrm{prop}}\left\{ \p_t u^n+d_n u^n\right\}\rho^n \, ,
\ee
with $d_n=d-2n$ the canonical mass dimension of the dimensionful couplings $\bar u^n$. Projecting the RHS, Eq.~\eqref{RHS}, onto the operators \eqref{projbasis} is akin to Taylor-expanding this RHS around $\rho=0$,
\be
\label{omegafk}
I_0[\mcF_k](\rho)=\sum_n\omega^n(u)\rho^n \, ,
\ee
and
\be
\label{tildeomegafk}
\left(k^{-d+2}\frac{\mrm d}{\mrm dt} f_k'\right)(\rho)I_1[\mcF_k](\rho)+\left(k^{-d+4}\frac{\mrm d}{\mrm dt} f_k''\right)(\rho)I_2[\mcF_k](\rho)=\sum_{n,m}\tilde\omega^n{}_m(u)\left[\p_t u^m+d_m u^m\right]\rho^n \, .
\ee
Since the LHS of Eq.\ \eqref{tildeomegafk} is independent of $\bar{u}^0$, we have $\tilde\omega^n{}_0\equiv 0$. The structure highlighted on its RHS then arises from the expansions
\be\label{fderivatives}
\begin{split}
	\left(k^{-d+2}\frac{\mrm d}{\mrm dt} f_k'\right)(\rho)&=\sum_{n=0}n[\p_t u^n+d_n u^n]\rho^{n-1} \, ,\\
	\left(k^{-d+4}\frac{\mrm d}{\mrm dt} f_k''\right)(\rho)&=\sum_{n=0}n(n-1)[\p_t u^n+d_n u^n]\rho^{n-2} \, .
\end{split}
\ee
In this way, the definitions of the vector $\omega^n(u)$ and the matrix $\tilde\omega^n{}_m(u)$ agree precisely with the definitions given in Eq.~\eqref{LHSRHSTMWetterich}. The explicit form for $\omega^n(u)$ and $\tilde\omega^n{}_m(u)$ are readily obtained using Computer-Algebra-Sytems. 

Combining these results with Eqs.~\eqref{master2} and~\eqref{master3}, it is straightforward to obtain fixed points $\{u^n_*\}$ as well as the stability matrix $B_*^{\rm FRGE}$. In the present work, for the case $d=4$, we will employ the particular fixed point values $\{u^n_*\}$ obtained in \cite{Codello:2008vh}. For $N_\mrm{prop}=8$ these are given by (see \cite[Table III]{Codello:2008vh}):
\bg\label{eq:B3.3}
(u^0_*,u^1_*,\dots,u^8_*)=10^{-3}(5.066,-20.748,0.088,-8.581,-8.926,-6.808,1.165,6.196,4.695) \, .
\eg
When using fixed point values for propagators with $N_\mrm{prop}<8$, we will simply cut off the list \gl{B3.3} at the corresponding value instead of using the fixed point values obtained from the $N_\mrm{prop}<8$-truncation. We do so because the list \gl{B3.3} represents a more accurate approximation of the fixed point position $\{u_*\}$ also at lower orders of the truncation.

In order to compare these results with the scaling corrections obtained from $\gamma_m{}^n$ for the family of composite operators $\{\cO_n=\int\dd x\sgo R^n\ |\ n=0,1,\dots,N_\mrm{scal}\}$, we must evaluate Eq.~\eqref{compFRGE2}. In this case only the TT- and trace-modes contribute, 
 \bg
 \label{eq:B4.1}
 \spl{
\sum_m\gamma_n{}^m(u)\rho^m =:&\,I_1+I_2 \, ,
 }
 \eg
where
\be
\begin{split}
   I_1 := & -\frac{k^{d_n}}{2\tilde V}\Tr_\mrm{TT}\bigg[\lef(\frac{\mrm d}{\mrm dt}\cR_{k,\mrm{TT}}\ri)
\lef({\bar\Gamma_{k,\mrm{TT}}}^{(2)}+{\cR_{k,\mrm{TT}}}\ri)^{-1}{\cO_n^{(2)}}{}_{\mrm{TT}}\lef({\bar\Gamma_{k,\mrm{TT}}}^{(2)}+\cR_{k,\mrm{TT}}\ri)^{-1}\bigg]  \, , \\
   I_2 := & -\frac{k^{d_n}}{2\tilde V}\Tr_\mrm{S}\bigg[\lef(\frac{\mrm d}{\mrm dt}\cR_{k,\hat h\hat h}\ri){}\lef({\bar\Gamma_{k,\hat h\hat h}}^{(2)}+\cR_{k,\hat h\hat h}\ri)^{-1}{\cO_n^{(2)}}{}_{\hat h\hat h}\lef({\bar\Gamma_{k,\hat h\hat h}}^{(2)}+\cR_{k,\hat h\hat h}\ri)^{-1}\bigg] \, . \\
\end{split}
\ee
 
 In order to read off the anomalous-dimension matrix $\gamma_n{}^m$, we have to calculate the two traces $I_1$ and $I_2$ that built the RHS of \Gl{B4.1}, and Taylor-expand them in powers of $\rho$. For the propagators arising from the Einstein-Hilbert truncation, this computation has been carried out in \cite{Houthoff:2020zqy,Kurov:2020csd}. In order to be able to compare to the polynomial $f(R)$-truncations of the Wetterich equation, we extend this study to the propagators \eqref{Gamma2} retaining terms in the expansion \eqref{ansatzfR} up to $N_{\rm prop}$. Since this particular  calculation has not been performed in the literature before, we have collected the details in Appendix~\ref{App.A}. Note that the composite operator equation~\gl{B4.1} allows us to truncate the operator bases $\{\cO_n=\int\dd x\sgo R^n, n=0,\cdots,N_{\rm scal}\}$ and the propagator at different values. These are then denoted by $N_\mrm{scal}$ and $N_\mrm{prop}$, respectively. The FRGE formalism above obviously only admits the case $N_\mrm{scal}=N_\mrm{prop}$. As for the case of the stability matrix $B_*^{\rm FRGE}$, results for $\gamma_n{}^m$ are readily obtained using Computer-Algebra-Software.   
 
\begin{figure}[t!]
\centering
  \includegraphics[width=0.49\textwidth]{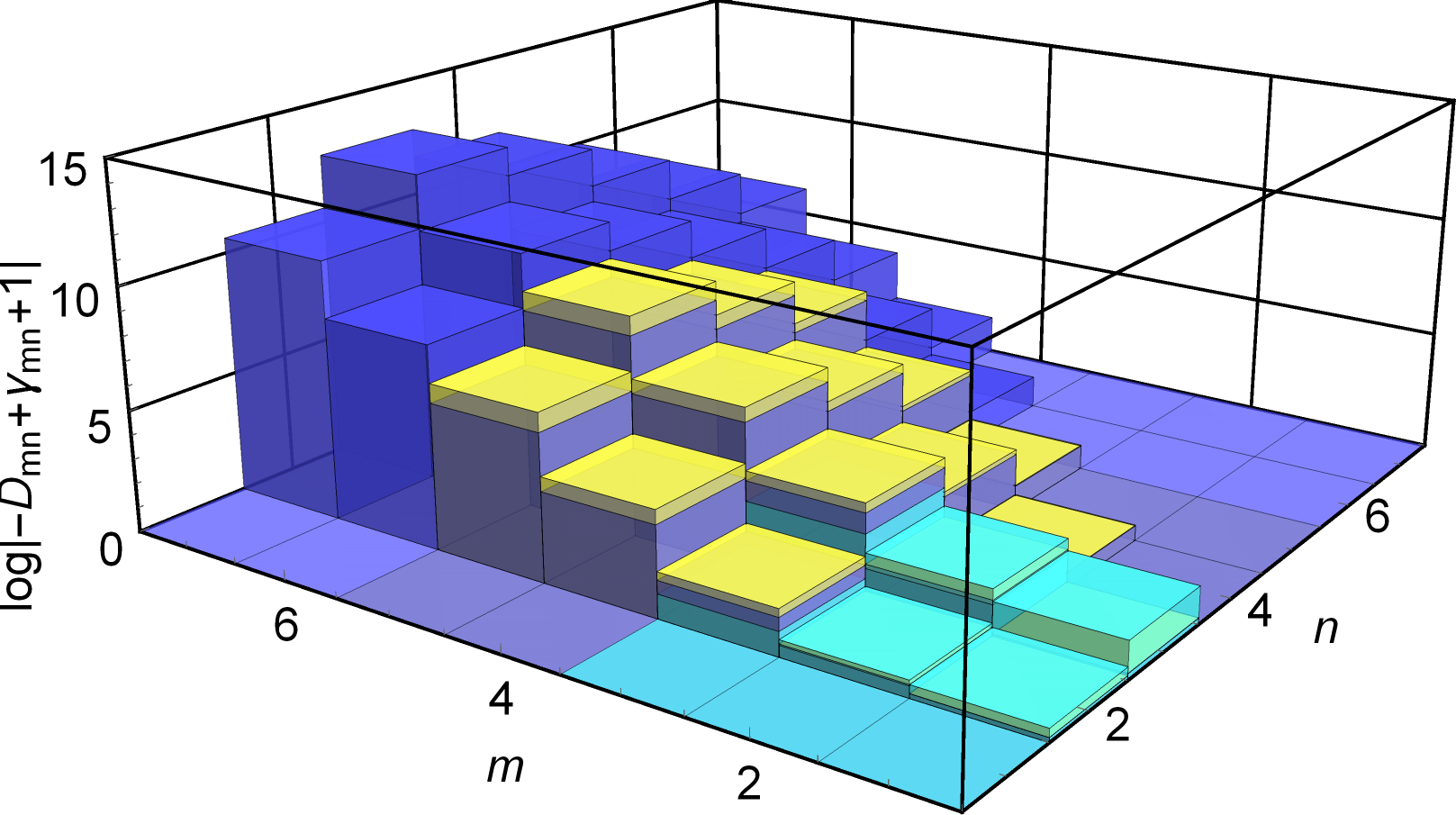} \,
  \includegraphics[width=0.49\textwidth]{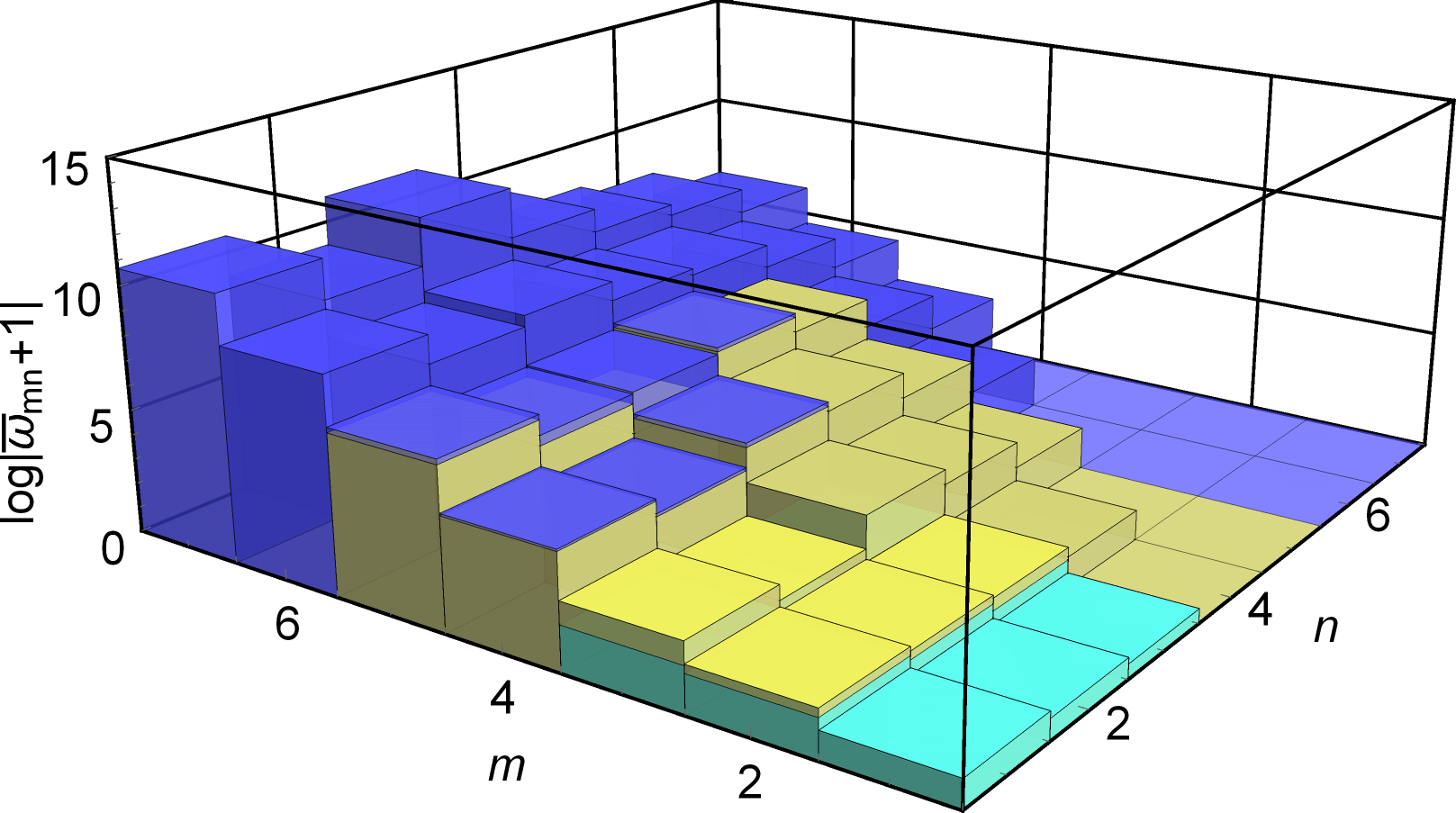} \\
  \includegraphics[width=0.49\textwidth]{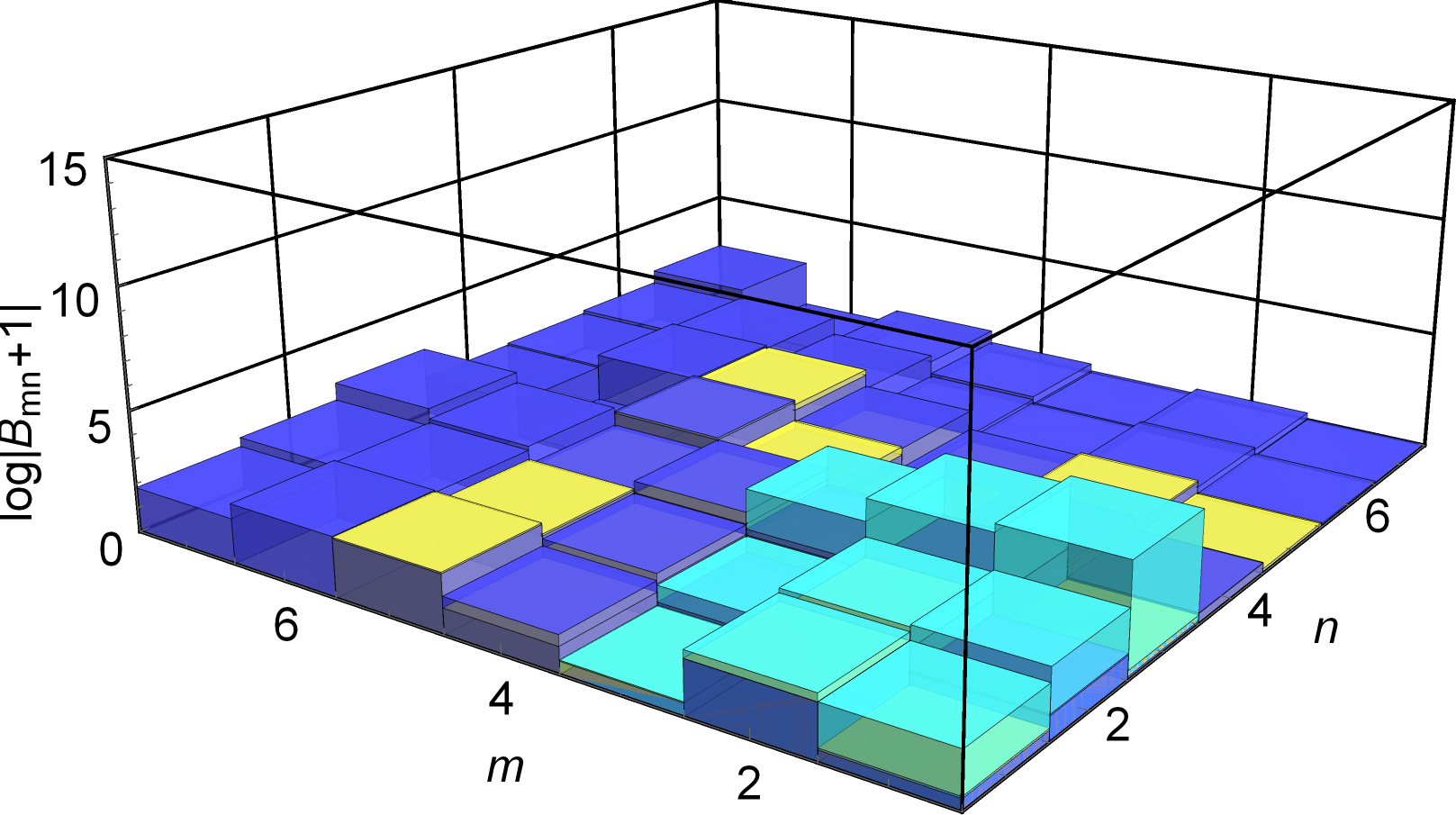}
\caption{\label{fig.1} Illustration of the size of the matrix entries appearing in $-D+\gamma^\mrm{T}(u_*)$ (upper left), $\tilde{\omega}(u_*)$ (upper right) and the stability matrix $B_*$ (lower row) entering the master equation \eqref{master3} on a logarithmic scale. The colors refer to the systems with $N_{\rm scal} = N_{\rm prop} = 3$ (cyan), $N_{\rm scal} = N_{\rm prop} = 5$ (yellow), and $N_{\rm scal} = N_{\rm prop} = 7$ (blue). The top lines illustrate that the size of the entries grow with the matrix indices $m,n$. At the level of the stability matrix, it is the interplay between the two contributions in the top row, which significantly reduces the size of the coefficients appearing in Eq.\ \eqref{Bfrge}.}
\end{figure}

 At this point, we are ready to compare the stability matrices \eqref{Bfrge} and \eqref{Bapproxgamma2}.  For concreteness, we limit the discussion to $d=4$. Inspecting Eq.~\eqref{Bapproxgamma2}, one might naively infer from here that the spectrum of $B^\mrm{COE}_*$ is quite similar to that of $B^\mrm{FRGE}_*$, with the anomalous scaling dimensions $\gamma_m{}^n(u_*)$ providing corrections to $-D^n_m$ that become smaller and smaller as $N_\mrm{prop}=N_\mrm{scal}\to\infty$.
  In Fig.\ \ref{fig.1}, we illustrate the entries of their building blocks $-D+\gamma^{\rm T}$ (top left diagram), $\tilde{\omega}$ (top right diagram) and $B_*$ (bottom) for various values of $N_\mrm{scal}$ and $N_\mrm{prop}$ \emph{on a logarithmic scale}.
 This decomposition actually establishes that the opposite is true: The entries of $\gamma_m{}^n(u_*)$ grow exponentially as $N_\mrm{prop}=N_\mrm{scal}\to\infty$. In addition, Fig.\ \ref{fig.1} also depicts that the proliferation of matrix entries is not limited to $\gamma_m{}^n(u_*)$, but also happens in an almost identical fashion for $\tilde\omega^n{}_m(u_*)$. It is striking to observe how in Eq.~\eqref{master3} these two rampant, and seemingly unphysical, matrices combine to $B^\mrm{FRGE}_*$ whose entries' magnitudes are in perfect unison with $-D^n_m$.

As a consequence of the exponential growths in the quantum parts, the spectra of $B^\mrm{COE}_*$ and $B^\mrm{FRGE}_*$ disagree even at a qualitative level, cf. Table~\ref{tab}. Qualitative agreement -- in terms of an agreement on the sign of the critical exponents -- is only assured for small-size truncations up to $N_\mrm{prop}=N_\mrm{scal}=2$. For larger-size truncations, the entries of $\gamma_m{}^n(u_*)$ reach a size that distorts the spectrum of $B^\mrm{COE}_*$ altogether, with the quantum corrections exceeding the classical (Gaussian) scaling dimensions by orders of magnitude. In Table~\ref{tab} this becomes evidently clear for $N_\mrm{prop}=N_\mrm{scal}=6$. These corrections surely seem intuitively anomalous and far from a physically reliable estimate. This is a paradoxical situation since, as we had mentioned earlier on, there exist plenty of physical systems, e.g. in the realm of condensed matter physics, for which the critical exponents emanating from $B^\mrm{COE}_*$ constitute actual physical observables.

As regards to observables, we would like to point out another striking results that Table~\ref{tab} exhibits.  For geometrical operators depending on some characteristic scale, the (negative) eigenvalues of $-D+\gamma$ can also be interpreted as the geometric scaling dimensions of these operators. In case of a single-operator approximation given by the spacetime volume $\cO_0$ (i.e., $N_\mrm{scal}=0$), we find dimensional reduction of its scaling dimension from classically 4 down to $2.04$. Albeit a gauge-dependent result, this strikingly confirms the dimensional reduction of spacetime in the UV from 4 to 2 dimensions that has already been observed in a variety of settings \cite{Carlip:2019onx}, including Asymptotic Safety \cite{Reuter:2012xf} and CDT \cite{Loll:2019rdj}.

\begin{table}[t!]
\centering
\begin{tabular}{|cc|c|ccccccc|}
\hline
$N_{\rm scal}$ & $N_{\rm prop}$ &  & $\lambda_0$
& $\lambda_1$ & $\lambda_2$ & $\lambda_3$ & $\lambda_4$ & $\lambda_5$ & $\lambda_6$
\\ 
\hline
&& $-D$ & $-4$ & $-2$ & $0$ & $2$ & $4$ & $6$ & $8$ 
\\ 
\hline
0 & $\geq 3$ & $B_*^{\rm COE}$ & $-2.04$ & & & & & & 
\\ 
\hline
\multirow{3}{*}{$1$} & \multirow{3}{*}{$1$} & $\Big. B_*^{\rm FRGE}$ &  & $-2.09$ & & & & & \\ 
&& $\Big. B_*^{\rm COE}$ && $-1.91$ & & & & & \\ 
& & $\Big. B_*^{\rm COE+GF}$ && $-1.94$ & & & & & \\
\hline
\multirow{2}{*}{$2$} & \multirow{2}{*}{$2$} & $\Big. B_*^{\rm FRGE}$ & $-27.02$ & \multicolumn{2}{c}{$-1.26\pm 2.45\I$} & & & & \\ & &  $\Big. B_*^{\rm COE}$ & \multicolumn{2}{c}{$-4.06\pm 1.38\I$} & $-0.89$  & & & &  \\
\hline
\multirow{2}{*}{$4$} & \multirow{2}{*}{$4$}& $\Big. B_*^{\rm FRGE}$ & 
\multicolumn{2}{c}{$-2.84 \pm 2.42\I$} & $-1.54$ & $4.27$ &  $5.09$ & & \\
&  & $\Big. B_*^{\rm COE}$ &  
\multicolumn{2}{c}{$-3.07\pm 2.79\I$} & $3.83$ & $49.1$ & $290.7$ & & \\
\hline
\multirow{3}{*}{$6$} & \multirow{3}{*}{$6$} & $\Big. B_*^{\rm FRGE}$ & 
\multicolumn{2}{c}{$-2.39 \pm 2.38\I$} & $-1.51$ & $4.16$ & \multicolumn{2}{c}{$4.68 \pm 6.09\I$} &  $8.68$    \\ 
 &  & $\Big. B_*^{\rm COE}$ & \multicolumn{2}{c}{$-21.30 \pm 22.10\I$} & $-4.53$ & \multicolumn{2}{c}{$0.66 \pm 2.35\I$}& $329.0$ & $954.9$ \\
 & & $\Big. B_*^{\rm loc}$ &  & 
 $-(26 + 4.4   d_1) 10^{10}$ &  $-6$  & $-4$ & $-2$ & $0$ & $2$ 
\\
\hline
\multirow{2}{*}{$6$} & \multirow{2}{*}{$2$} & $\Big. B_*^{\rm FRGE}$ & \multicolumn{2}{c}{-} & - & 
 \multicolumn{2}{c}{-} & -  & - \\
 & & $\Big. B_*^{\rm COE}$ & \multicolumn{2}{c}{$-11.74 \pm 19.49\I$} & $-9.80$ & 
 \multicolumn{2}{c}{$-2.41 \pm 1.88\I$} & $-0.15$ & $3.64$ \\
\hline
\end{tabular}
\caption{\label{tab} Illustration of the spectral properties of the matrices $B_*^{\rm FRGE}$ and $B_*^{\rm COE}$ for several values of $N_{\rm scal}$ and $N_{\rm prop}$. The first line (classical scaling) gives the entries of $-D$. Complex eigenvalues are presented pairwise in a double-column entry. The case $N_{\rm scal} = N_{\rm prop} = 1$ summarizes the results of Sect.\ \ref{sec22} while the lower entries compare the spectra of the stability matrix obtained from solving the FRGE and COE for polynomial $f(R)$-truncations. Note that the definition of $B_*^{\rm FRGE}$ entails that $N_{\rm scal} = N_{\rm prop}$. The results from the ``localization approximation'' (loc) are referenced from Eq.\ \eqref{Bmatloc}.}
\end{table}

\section{Almost-Gaussian Scaling from the Quantum Corrections}
\label{sect.analytic}
The take-away message from Fig.\ \ref{fig.1} is that the spectrum of the stability matrix is essentially determined by the interplay between the quantum contributions to Eq.\ \eqref{Bfrge}. At this point, it is highly interesting to understand the mechanism that actually underlies our findings in Table \ref{tab}.
The key ingredient is the (regularized) two-point correlation function $(\Gamma_k^{(2)} + \cR_k)$. In general, this operator is matrix-valued in field space. A detailed inspection of this matrix shows, that it is only one specific matrix element which drives the mechanism, the one associated with the trace part $\hat{h}$ appearing in the decomposition \eqref{TTdec}. 

\begin{figure}[t]
\includegraphics[width=0.5\textwidth]{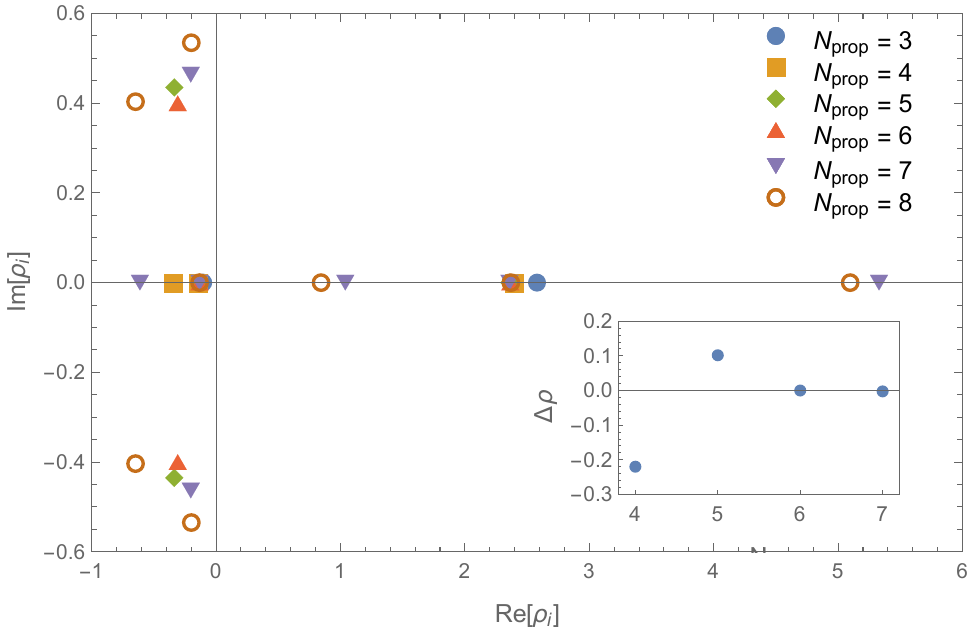} \,
\includegraphics[width=0.455\textwidth]{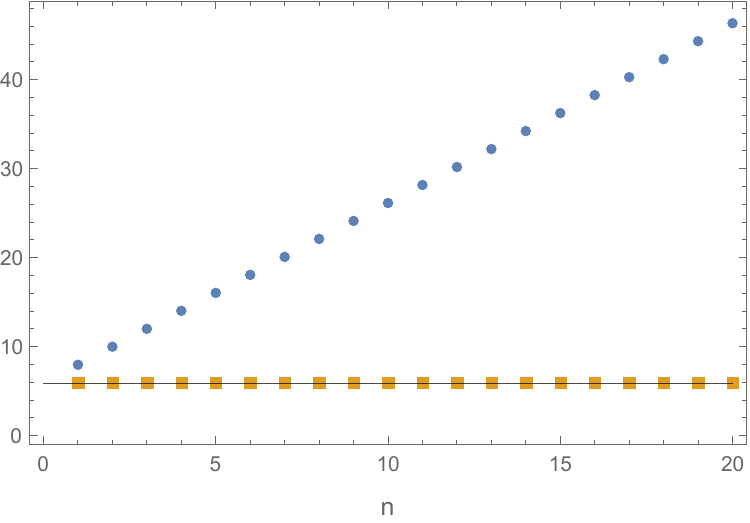}
\caption{\label{fig.2} Illustration of the analytic structure of $(\Gamma_{k,\hat{h}\hat{h}}^{(2)} + \cR_{k,\hat{h}\hat{h}})$ (left) and the coefficients appearing in the series expansion eq.\ \eqref{polyexp} (right).}
\end{figure}

The first step is to understand the root structure of \eqref{Gamma2}. For this purpose, we restrict to $d=4$, substitute the expansions of $f(\Rb)$ up to $N_{\rm prop} = \{3,4,5,6,7,8\}$, and convert to dimensionless variables. Substituting the fixed point position \eqref{eq:B3.3}, this leads to a polynomial of order $N_{\rm prop}$ in the dimensionless curvature $\rho$. For $N_{\rm prop} = 8$, the resulting expression is given by
\be\label{roots1}
\begin{split}
(\Gamma_{k,\hat{h}\hat{h}}^{(2)} + \cR_{k,\hat{h}\hat{h}})(u_*) = \frac{k^4}{4} & \, \big(0.197 \rho ^8-1.279 \rho ^7+0.958 \rho ^6+2.072 \rho ^5+0.976 \rho ^4
\\ & \, \, 
-0.707 \rho ^3   -0.732 \rho ^2-0.464 \rho -0.051\big) \, . 
\end{split}
\ee
The roots of these polynomials are easily found numerically and depicted in the left panel of Fig.\ \ref{fig.2}. Increasing $N_{\rm prop}$ adds a new root in each step. The most important one is the one which is closest to the expansion point $\rho = 0$. This root appears at $N_{\rm prop} = 3$ and remains stable when $N_{\rm prop}$ is increased. For $N_{\rm prop} = 8$, its position is given by
\be\label{scalarroot}
\rho_0^{N_{\rm prop} = 8} = -0.133 \, . 
\ee
The relative distance between the root found at lower orders, $\Delta \rho \equiv \frac{\rho_0^{N_{\rm prop} = n}-\rho_0^{N_{\rm prop} = 8}}{\rho_0^{N_{\rm prop} = 8}}$, is displayed in the inset, demonstrating the stability of the root with respect to in increase of $N_{\rm prop}$. The corresponding analysis in the transverse-traceless sector shows that in this case the root closest to the origin is at $|\rho_{\hat h \hat h}^{N_{\rm prop} = 8}| \approx 0.66$.

It turns out that it is the root \eqref{scalarroot} that dominates the expansion of the flow equations in the dimensionless Ricci scalar $\rho$. This is seen as follows. The Wetterich and composite operator equations contain the inverse of the two-point function. For the specific gauge choice the matrix $(\Gamma_k^{(2)} + \cR_k)$ diagonalizes, so the inverse contains the inverse of $(\Gamma_{k,\hat{h}\hat{h}}^{(2)} + \cR_{k,\hat{h}\hat{h}})$ which is expanded at $\rho = 0$. Evaluating this expansion at the fixed point $u_*$ yields
\be\label{polyexp}
(\Gamma_{k,\hat{h}\hat{h}}^{(2)} + \cR_{k,\hat{h}\hat{h}})^{-1}(u_*) = \sum_n \tau^n(u_*) \rho^n \, .
\ee
The expansion coefficients $\tau^n(u_*)$ are illustrated in the right panel of Fig.\ \ref{fig.2}. Here the blue dots correspond to $\log|\tau^n(u_*)|$ while the orange plots display $\log|\tau^n(u_*) (\rho_0^{N_{\rm prop} = 8})^n|$. Owed to the pole in the complex unit circle, the coefficients grow exponentially. This is illustrated by the blue dots. The orange points show that the growths is controlled by $\rho_0^{N_{\rm prop}=8}$ already for low values of $n$. Thus, we recover the familiar result that the growths of the expansion coefficients is controlled by the root closest to the expansion point\footnote{A simple example illustrating this effect is the expansion ${1}/({1-10x})=\sum_n 10^n\,x^n$ where the expansion coefficients are determined by the pole at $x = 1/10$.}
\be
\lim_{n \rightarrow \infty} \, \left| \frac{\tau^{n+1}(u_*)}{\tau^{n}(u_*)} \right| = \frac{1}{\rho_0^{N_{\rm prop}=8}} \, . 
\ee
As shown in the right diagram of Fig.\ \ref{fig.2}, this relation is already fulfilled to very good accuracy at low values $n \simeq 10$. It is straightforward to repeat this analysis for the other matrix elements appearing in $(\Gamma_k^{(2)} + \cR_k)$. Since the root in the scalar sector is the one closest to the expansion point, the expansion coefficients in this sector exhibit the fastest growth and therefore determine the structure of the flow equation at large $N_{\rm prop}$. This feature propagates into the entries of $\omega^n(u_*)$ and $\tilde\omega^n{}_m(u_*)$ and also to $\gamma_m{}^n(u_*)$ and $\tilde\gamma_m{}^n(u_*)$.

The exponential growth of the expansion coefficients \eqref{polyexp} entails that it is the \emph{scalar trace which dominates} the beta functions for couplings $\bar{u}^n$ with $n$ sufficiently large. In particular, the classical contributions originating from the LHS of the flow equation (encoded in $D$) are negligible. This suggests that for sufficiently large values of $n$, a good approximation of the flow is given by
\be\label{localization1}
\frac{1}{2} {\rm Tr}_S \bigg[\lef({\Gamma_{k,\hat h\hat h}}^{(2)}+\cR_{k,\hat h\hat h}\ri)^{-1} \, \lef(\frac{\mrm d}{\mrm d t} \cR_{k,\hat h\hat h}\ri) \bigg] \approx 0 \, . 
\ee
We will call this the ``localization'' (loc) approximation, since the analytic structure of the propagator essentially localizes the contributions of the RG flow in one single sector. 

Since \eqref{localization1} is a pure quantum expression, this immediately raises the puzzle about how the localization approximation may recover any knowledge about the classical scaling dimensions $d_{u^n}$. In order to answer this question, let us evaluate the RG-time derivative of \eqref{cRhh},
\be\label{dtRk}
\begin{split}
\frac{\mrm d}{\mrm d t}\cR_{k,\hat h\hat h} =  \frac{3}{4} & \Big\{ \left[f^\prime_k + 6 f^{\prime\prime}_k (\Delta + R_k) - 2 f^{\prime\prime}_k \Rb \right] \p_t R_k \\
& \; + \left[ (\mrm d f^\prime_k /\mrm d t) + 3 (2 \Delta + R_k) (\mrm d f^{\prime\prime}_k/\mrm d t) - 2 \Rb (\mrm d f^{\prime\prime}_k/\mrm d t )  \right] R_k \Big\} \, ,
\end{split}
\ee
where the $t$-derivatives of $f_k(R)$ where given in Eq.\ \eqref{fderivatives}. The first line in \eqref{dtRk} will give the contributions to $\omega^n(u_*)$ while the second encodes the terms appearing in $\tilde{\omega}^n{}_m(u_*)$. Moreover, Eq.\ \eqref{fderivatives} shows that \eqref{localization1} actually gives beta functions for the couplings $u^n$, $1 \le n \le N_{\rm prop}$.

We can then evaluate the trace \eqref{localization1} using the zeroth order in the heat kernel and approximating the propagator by its expression at the fixed point $u_*$. The approximation in the heat-kernel expansion  is again justified by the exponential growth of the series coefficients $\tau^n$ to which higher-order contributions of the heat-kernel are subleading. Solving for $\p_t u^n$, this results in beta functions which are linear in the couplings $u^n$. The stability matrix is then obtained by taking the derivative with respect to these couplings. This results in the upper-triangular matrix. Thus the eigenvalues of $B_*^{\rm loc}$ are given by its diagonal entries. For example, $N_{\rm prop} = 6$ leads to 
\be\label{Bmatloc}
 B_*^{\rm loc} = 
 \left(
\begin{array}{cccccc}
 (-4.4\times 10^{10} d_1-2.6\times 10^{11}) & -3.68\times 10^{16} & -243. & -1458. & -6561. & -2.53 \times 10^4 \\
\Big. 0 & -d_2-6 & 27 & 162 & 729 & \frac{19683}{7} \\
\Big. 0 & 0 & -d_3-6 & 12 & 54 & \frac{1458}{7} \\
\Big. 0 & 0 & 0 & -d_4-6 & 9 & \frac{243}{7} \\
\Big. 0 & 0 & 0 & 0 & -d_5-6 & \frac{54}{7} \\
\Big. 0 & 0 & 0 & 0 & 0 & -d_6-6 \\
\end{array}
\right)
\ee
 Notably, these contain the classical scaling dimension of the couplings. Thus one can indeed recover an almost classical scaling from \eqref{localization1}. The specific structure of the regulator \eqref{cRhh} is crucial for that, however.

This result also clarifies why the COE does not recover the almost-Gaussian scaling behavior. The localization mechanism owed to the pole in the propagator is still operational in this case. However, the regulator is no longer tied to the operators $\cO_n$. In particular $\cR_k$ is independent of the anomalous scaling dimension. Thus, the mechanism leading to stability matrices of the form \eqref{Bmatloc} fails in this case.

\section{Discussion and Conclusions}
\label{sec:conclusion}
Interacting renormalization group (RG) fixed points provide an attractive mechanism for providing a consistent and predictive high-energy completion of a theory beyond the realm of perturbation theory. The predictive power of such a fixed point is encoded in the spectrum of the stability matrix $B_*$, Eq.\ \eqref{stabmat}, controlling the linearized RG flow in its vicinity. The master equation derived in 
Eq.\ \eqref{master3} encodes the general structure of this matrix separating the contributions originating from the classical scaling, quantum corrections, and regulator piece. This result is general in the sense that it holds for any fixed point irrespective of a given field content. Results obtained from the Wetterich equation and the composite operator equation can readily be cast into this parameterization. This highlights the structural differences occurring when evaluating the operator spectrum within these distinct computation schemes.

We then applied the general formalism to the Reuter fixed point projected on $f(R)$-type interactions polynomial in the Ricci scalar $R$. The evaluation of the stability matrix based on the Wetterich equation yields a spectrum of $B_*$ which almost follows the canonical scaling for large powers of $R$. This is a key result supporting the predictive power of the Reuter fixed point. In the composite operator equation, this feature is absent though. Our general framework provides a comprehensive understanding of this mismatch in the following way: near the Reuter fixed point the regularized two-point function associated with the trace sector in the gravitational fluctuations exhibits a pole close to the expansion point of the functional renormalization group equations. This pole dominates the structure of the stability matrix for polynomials of sufficiently high order in $R$. The flow equations are well-approximated by the quantum contributions stemming from the trace-modes and the classical part of the stability matrix does not play any role. This mechanism is operative for both the Wetterich and the composite operator equation. The difference in the spectra of $B_*$ can then be traced to the different regulator prescriptions. The regularization based on the Wetterich equation is constructed in such a way that also the dominant quantum contributions to the stability matrix exhibit classical scaling properties. For the composite operator equation this feature of the regulator and hence the almost-Gaussian scaling of the spectrum of $B$ is absent. We stress that low-order expansions are unaffected by this mechanism though and shows qualitative agreement within the two complementary approaches.

Similar observations about the regulator dependence of the spectrum have recently been made in the context of functional $f(R)$-truncations using non-adaptive regularization schemes \cite{Mitchell:2021qjr,Morris:2022btf}. Studying the spectrum of admissible perturbations at a fixed functional $f_*(R)$ using Sturm-Liouville theory, it is found that the spectrum of $B_*$ may either come with a finite or infinite number of relevant eigendirections, depending on the precise implementation of the regulator in the trace sector
\cite{Mitchell:2021qjr}.

These findings raise the crucial question whether the almost-Gaussian scaling is indeed a genuine feature of the Reuter fixed point or a build-in feature of the key equation used to study it. One way to shed light on this question would be the study of $f(R)$-type projections using an expansion based on an operator basis of the form $\{\cO_n=\int\dd x\sgo(R-R_0)^n\}$, where the reference curvature $R_0$ is chosen in such a way that there are no propagator poles within the unit circle centered at the expansion point. Alternatively, a detailed understanding of the poles contained in $(\Gamma_k^{(2)} + \mcR_k)^{-1}$ on a generic, curved background spacetime may also give further information about the predictive power of the gravitational asymptotic safety program. This assessment also connects to the realm of functional $f(R)$-truncations where the pole structure of $(\Gamma_k^{(2)} + \cR_k)^{-1}$ also played a crucial role in obtaining isolated, global fixed functionals $f_k(R)$ by insisting that the solutions can pass through the poles of the flow equation located on the real axis \cite{Dietz:2012ic,Demmel:2012ub}.

We note that the mechanism uncovered in this article may actually not be limited to the realm of quantum gravity. The interplay between poles of a two-point function and the spectrum of the stability matrix may be realized in other settings as well. Thus, the effect of quantum-induced almost-Gaussian scaling may occur much more frequently. Clearly, a general understanding of this situation would be highly desirable also beyond the realm of $f(R)$-computations and we hope to come back to this in the future.

\noindent
\subsection*{Acknowledgements}

\noindent
We thank Martin Reuter, Carlo Pagani, Luca Buoninfante, and Rudrajit Banerjee for interesting discussions. MB gratefully acknowledges financial support by the Deutsche Forschungsgemeinschaft (DFG, German Research Foundation) -- project number 493330310. The work of AK on the spectrum of the stability matrix is supported by the Russian Science Foundation grant No 23-12-00051, https://rscf.ru/en/project/23-12-00051/.

\appendix
\section{Calculation of $\gamma_n{}^m$ for the operator family $\{\int\dd x\sgo R^n\}$}
\label{App.A}
This appendix collects the technical details entering into the evaluation of the operator traces in Eq.~\gl{B4.1}. The computation utilizes the heat-kernel technology reviewed in \cite{Codello:2008vh} and we outline the relevant steps in order to render the present work self-contained.

Inspecting the traces in Eq.~\gl{B4.1}, we see that their arguments are given by operator-valued functions $W_s(\Delta)$ where $\Delta := - \gb^{\mu\nu} \Db_\mu \Db_\nu$ is the Laplacian constructed from the background metric and the trace is either over scalars ($s=S$) or transverse-traceless symmetric matrices ($s = {\rm TT})$. Our computation tracks powers of the background scalar curvature $\Rb$ only. Hence it suffices to work on a spherically symmetric background. 

We then evaluate the traces using the early-time expansion of the heat kernel. Exploiting that we work on a background sphere, this expansion takes the form
\bg
\label{eq:B6.1}
\Tr_{s}\lef[W(\Delta)\ri]=\frac{1}{(4\pi)^{d/2}} \sum_{m=0}^\infty \alpha_m^{s} \, Q_{\frac{d}{2}-m}[W] \, \int\dd x\sgbo  \, \Rb^m  \, .
\eg
Here $\alpha_m^s$ are the deWitt coefficients with $\Rb = 1$. For the background sphere these numbers can be found in \cite{Kluth:2019vkg} and we collect the relevant ones in Table \ref{Tab.2}.
\begin{table}[t!]
\centering
\begin{tabular}{|c|c|c|c|c|c|c|c|}
\hline
\Bigg.  $m$ & 0 & $1$ & $2$ & $3$ & $4$ & $5$ & $6$ \\ \hline
\Bigg. $\alpha_m^S$ & $1$ & $\frac{1}{6}$ & $\frac{29}{2160}$ & $\frac{37}{54432}$ & $\frac{149}{6531840}$ & $\frac{179}{431101440}$ & $-\frac{1387}{201755473920}$ \\
\Bigg. $\alpha_m^{\rm TT}$ &
$5$ & $-\frac{5}{6}$ & $-\frac{1}{432}$ & $\frac{311}{54432}$ & $\frac{109}{1306368}$ & 
& 
\\ \hline 
\end{tabular}
\caption{\label{Tab.2} Summary of the relevant deWitt coefficients for the Laplacian on a four-dimensional background sphere \cite{Kluth:2019vkg}. The coefficients not listed in the table do not enter into our computation.}
\end{table}
The $Q$-functionals are given in terms of the Mellin transform of the trace argument. For positive index $m > 0$,
\bg
\label{eq:B6.2}
Q_m[W]=\frac{1}{\Gamma(m)}\int_0^\infty\D z\,z^{m-1}W(z) \, .
\eg
For $m \le 0$ one can integrate by parts to obtain
\be
Q_{-m}[W] = \frac{(-1)^k}{\Gamma(m+k)} \int_0^\infty\D z\,z^{m+k-1}W^{(k)}(z) \, , \quad m + k > 0 \, . 
\ee
For $m+k$ being integer, this reduces to the relation $Q_{-m}[W] = (-1)^m \, W^{(m)}(0)$. Based on Eq.\ \eqref{eq:B6.2} one furthermore establishes the identity that 
\bg
\label{eq:B6.3}
Q_m\lef[z^p \, W\ri]=\frac{\Gamma(m+p)}{\Gamma(m)}Q_{m+p}[W] \, .
\eg
In order to transition to dimensionless quantities, we finally define
\bg
\label{eq:B7.6}
\spl{
q_m^\mrm{S}&:=(k^2)^{-m-2}Q_m[W^\mrm{S}(\,\cdot\,;k^2\rho)] \, ,\\
q_m^\mrm{TT}&:=(k^2)^{-m-1}Q_m[W^\mrm{TT}(\,\cdot\,;k^2\rho)] \, .
}
\eg

Based on these prerequisites, we proceed by evaluating $I_1$ and $I_2$. The composite operator Hessian $\cO^{(2)}_n$ can be found in~\cite{Kurov:2020csd} and the propagator modalities are as in~\cite{Codello:2008vh}. This results in the trace arguments
\bg
\label{eq:B4.8}
\begin{split}
W^\mrm{TT}(\Delta;\Rb)= & \, \lef[-\eta_{f'_k(\Rb)}+\p_t\ri]R_k \, , \\
W^\mrm{S}(\Delta;\Rb)
= & \, 
4\lef(\p_t R_k+\zeta_{f_k(\Rb)}R_k\ri)\lef[(d-1)\Delta -\Rb\ri]  \\ & +2(d-1)R_k\lef[2\p_t R_k+\zeta_{f_k(\Rb)}R_k\ri]
+(d-2)\frac{f'_k(\Rb)}{f''_k(\Rb)}\lef[\p_t R_k-\eta_{f_k(\Rb)}R_k\ri] \, .
\end{split}
\eg
Here we specialized to the Litim-type cutoff \eqref{Rk-litim} and  introduced the anomalous dimensions
\bg
\label{eq:B4.7}
\begin{split}
\eta_{f_k(R)}&=-\frac{1}{f'_k(R)}\p_t f'_k(R)=-(d-2)-\frac{(\p_t\mathcal F_k')(\rho)}{\mathcal F_k'(\rho)}+2\rho\frac{\mathcal F_k''(\rho)}{\mathcal F_k'(\rho)} \, , \\
\zeta_{f_k(R)}&=\frac{1}{f''_k(R)}\p_t f''_k(R)=(d-4)+\frac{(\p_t\mathcal F_k'')(\rho)}{\mathcal F_k''(\rho)}-2\rho\frac{\mathcal F_k'''(\rho)}{\mathcal F_k''(\rho)} \, .
\end{split}
\eg

Using the operator insertion $\cO_n = \int \D^dx \sqrt{g} R^n$, the two traces in Eq.~\gl{B4.1} evaluate to
\bg
\label{eq:B7.8}
\spl{
I_1=&\frac{1}{(4\pi)^{d/2}}\mathcal F'_k(\rho)\rho^{n-1}\Bigg(\mathcal F'_k(\rho)\lef(1-\frac{2(d-2)}{d(d-1)}\rho\ri)+\mathcal F_k(\rho)\Bigg)^{-2}\\
&\times \Bigg\{\frac{d}{2}c_1^\mrm{TT}\alpha_0^\mrm{TT}q_{\frac{d}{2}+1}^\mrm{TT}+\sum_{m=0}^\infty\bigg(c_0^\mrm{TT}\rho\, \alpha_m^\mrm{TT}+c_1^\mrm{TT}(d/2-m-1)\alpha_{m+1}^\mrm{TT}\bigg)q_{\frac{d}{2}-m}^\mrm{TT}\Bigg\}
}
\eg
and
\bg
\label{eq:B7.9}
\spl{
I_2=\,&\frac{(1-d)}{(4\pi)^{d/2}}\,\mathcal F_k''(\rho)\rho^{n-2}\Bigg\{2(d-1)^2\mathcal F_k''(\rho)\lef[1-\frac{\rho}{d-1}\ri]^2\\
	&\quad\quad\quad\quad+(d-1)(d-2)\mathcal F_k'(\rho)\lef[1-\frac{\rho}{d-1}\ri]-(d-2)\rho \mathcal F_k'(\rho)+\frac{d(d-2)}{2}\mathcal F_k(\rho)\Bigg\}^{-2}\\
	&\times\Bigg\{\frac{d}{2}\lef[c_1^\mrm{S}\alpha_0^\mrm{S}\rho+\lef(\frac{d}{2}-1\ri)c_2^\mrm{S}\alpha_1^\mrm{S}\ri]q_{\frac{d}{2}+1}^\mrm{S}
		+\frac{d}{2}\lef(\frac{d}{2}+1\ri)c_2^\mrm{S}\alpha_0^\mrm{S}q_{\frac{d}{2}+2}^\mrm{S}\\
	&\quad\quad+\sum_{m=0}^\infty\Bigg[c_0^\mrm{S}\alpha_m^\mrm{S}\rho^2+\lef(\frac{d}{2}-m-1\ri)c_1^\mrm{S}\alpha_{m+1}^\mrm{S}\rho\\
	&\quad\quad\quad\quad\quad\quad+\lef(\frac{d}{2}-m-2\ri)\lef(\frac{d}{2}-m-1\ri)c_2^\mrm{S}\alpha_{m+2}^\mrm{S}\Bigg]q_{\frac{d}{2}-m}^\mrm{S}\Bigg\} \, .
}
\eg
The coefficients $c^s$ depend on $d$ and $n$ and are given by 
\be
\begin{split}
	c_0^\mrm{TT}=\frac{n(d-2)}{d(d-1)}-\foh\, ,\qquad &
    c_0^\mrm{S}=\frac{d(d-2)}{4}-{n(d-n-1)} \, ,  \\
   c_1^\mrm{TT}=-n/2 \, ,\qquad 
	& c_1^\mrm{S}=\frac{n(d^2-(4n-1)(d-1)-1)}{2}\, , \\
 & c_2^\mrm{S}={n(n-1)(d-1)^2} \, .
\end{split}
\ee

In principle, one can now expand Eqs.~\gl{B7.8} and \gl{B7.9} in powers of $\rho$ and subsequently read off $\gamma_{n}{}^m(u(k))$ according to \Gl{B4.1}. Owed to the $k$-derivatives acting on the couplings $u(k)$ in the second term of the anomalous dimensions \gl{B4.7} the resulting system is not self-contained though and requires an explicit expression for the beta functions $\p_t u^n(k) = \beta^n(u)$. At a fixed point $u^*$ the beta functions vanish by definition. This entails that 
\be
\mathcal F_k(\rho)\to\mathcal F_*(\rho)=\sum_{n=0}^{N_\mrm{prop}}u^n_*\rho^n \, , \quad (\p_t\mathcal F_*)(\rho)=0 \, . 
\ee
Substituting these conditions into the expressions for the anomalous dimensions gives
\bg
\label{eq:B8.1}
\spl{
\eta_{f_k(R)}\to\eta_*(\rho)&=-(d-2)+2\rho\frac{\mathcal F_*''(\rho)}{\mathcal F_*'(\rho)}\,,\\
\zeta_{f_k(R)}\to\zeta_*(\rho)&=(d-4)-2\rho\frac{\mathcal F_*'''(\rho)}{\mathcal F_*''(\rho)} \, .
}
\eg
By substituting the fixed point values $u_*$, given in \Gl{B3.3}, the results \eqref{eq:B7.8} and \eqref{eq:B7.9} allow to give the scaling dimension $\gamma_n{}^m(u_*)$ at the fixed point. 

\section{The flow equation for functional $f(R)$-truncations}
\label{App.B}
For completeness, we reproduce the functional f(R)-equation obtained in \cite{Machado:2007ea}.
\be\label{eq:B1} 
\begin{split}
& 384 \pi^2   \left( \p_t \cF_k + 4 \cF_k - 2 \rho \cF^{\prime}_k \right) = \vspace*{20cm} \\ 
& \quad \Big[ 5 \rho^2 \theta\left(1-\tfrac{\rho}{3}\right) 
-  \left( 12 + 4 \, \rho - \tfrac{61}{90} \, \rho^2 \right)\Big]
\Big[1 - \tfrac{\rho}{3} \Big]^{-1}  
+ 10 \, \rho^2 \, \theta\left(1 -\tfrac{\rho}{3}\right) \\
& + \Big[ 10 \, \rho^2 \, \theta(1-\tfrac{\rho}{4}) - \rho^2 \, \theta(1+\tfrac{\rho}{4}) 
-   \left( 36 + 6 \, \rho - \tfrac{67}{60} \, \rho^2 \right) \Big] 
\Big[ 1 - \tfrac{\rho}{4}\Big]^{-1} \\ 
& +  \Big[ \eta_f \, \left( 10 - 5  \rho - \tfrac{271}{36}  \rho^2 + \tfrac{7249}{4536}  \rho^3 \right) 
+ \left( 60 - 20  \rho - \tfrac{271}{18}  \rho^2 \right)
\Big] \left[ 1 + \tfrac{\cF_k}{\cF^\prime_k} - \tfrac{\rho}{3} \right]^{-1} \\ 
& + \frac{5\rho^2}{2} \, \Big[ 
 \eta_f \,
\left( (1+\tfrac{\rho}{3}) \theta(1+\tfrac{\rho}{3}) + (2+\tfrac{\rho}{3}) \theta(1+\tfrac{\rho}{6}) \right)
+ 2 \theta(1+\tfrac{\rho}{3}) + 4 \theta(1+\tfrac{\rho}{6}) 
\Big]  
 \left[ 1 + \tfrac{\cF_k}{\cF^\prime_k} - \tfrac{\rho}{3} \right]^{-1} \\ 
& + 
\Big[
\cF^{\prime}_k \, \eta_f \,
\left(6 + 3 \rho + \tfrac{29}{60} \rho^2 + \tfrac{37}{1512} \, \rho^3 \right)
+ \left( \p_t \cF^{\prime \prime}_k  - 2 \rho \cF^{\prime \prime \prime}_k \right) 
\left( 27 - \tfrac{91}{20} \rho^2  - \tfrac{29}{30} \, \rho^3 - \tfrac{181}{3360} \, \rho^4 \right) 
 \\ 
&  + \cF^{\prime \prime}_k \left( 216 - \tfrac{91}{5}  \rho^2 - \tfrac{29}{15}  \rho^3 \right)   
+  \cF^\prime_k \left( 36 + 12  \rho + \tfrac{29}{30}  \rho^2 \right)
\Big] \Big[ 2 \cF_k + 3 \cF^\prime_k (1-\tfrac{2}{3}\rho) + 9 \cF^{\prime \prime}_k (1-\tfrac{\rho}{3})^2 \Big]^{-1} \, , 
\end{split}
\ee
where we used the short-hand notation
\be
\eta_f = \frac{1}{\cF^\prime_k} \, \left( \p_t \cF^\prime_k +2 \cF^\prime_k - 2 \rho \cF^{\prime \prime}_k \right) \, .
\ee
This equation forms the basis for evaluating $B_*^{\rm FRGE}$ in Sec.\ \ref{sect32}.



\end{document}